\documentclass[letterpaper,twocolumn,10pt]{article}
\usepackage{usenix2019_v3}

\usepackage{tikz}
\usepackage{amsmath}

\usepackage{booktabs}
\usepackage{subfigure}
\usepackage{multirow}
\usepackage{xurl}

\usepackage{algorithmic}
\usepackage{algorithm} 
\usepackage[misc]{ifsym}
\usepackage[T1]{fontenc}

\usepackage{filecontents}

\begin{document}

\date{}

\pagestyle{empty}


\title{\Large \bf \textsc{SiamHAN}: IPv6 Address Correlation Attacks on TLS Encrypted Traffic via Siamese Heterogeneous Graph Attention Network}


\author{
{\rm Tianyu Cui$^{1,2}$, Gaopeng Gou$^{1,2}$, Gang Xiong$^{1,2}$, Zhen Li$^{1,2}$, Mingxin Cui$^{1,2}$, and Chang Liu$^{1,2,}$\thanks{Chang Liu is the corresponding author.}}\\
\\
{\rm $^1$Institute of Information Engineering, Chinese Academy of Sciences}\\
{\rm $^2$School of Cyber Security, University of Chinese Academy of Sciences}\\
{\rm \{cuitianyu, gougaopeng, xionggang, lizhen, cuimingxin, liuchang\}@iie.ac.cn}
}
\maketitle

\begin{abstract}
Unlike IPv4 addresses, which are typically masked by a NAT, IPv6 addresses could easily be correlated with users’ activity, endangering their privacy. Mitigations to address this privacy concern have been deployed, making existing approaches for address-to-user correlation unreliable. This work demonstrates that an adversary could still correlate IPv6 addresses with users accurately, even with these protection mechanisms. To do this, we propose an IPv6 address correlation model -- \textsc{SiamHAN}. The model uses a Siamese Heterogeneous Graph Attention Network to measure whether two IPv6 client addresses belong to the same user even if the user's traffic is protected by TLS encryption. Using a large real-world dataset, we show that, for the tasks of tracking target users and discovering unique users, the state-of-the-art techniques could achieve only 85\% and 60\% accuracy, respectively. However, \textsc{SiamHAN} exhibits 99\% and 88\% accuracy. 
\end{abstract}

\section{Introduction}
Over the past years, we have witnessed an increasing number of network providers expediting the deployment of IPv6~\cite{internetsociety2020world,SarrarMASU12}. Due to these efforts, one-third of Internet users can now access online services through IPv6 \cite{google2020ipv6}. With this growth in use, however, come increased focus on security and privacy issues (e.g.,~\cite{BergerKP20,LiF20,CzyzLAB16}). A particular concern for the privacy of IPv6 users is the user activity correlation attack~\cite{rfc7721}. In this attack, which works even on traffic encrypted by Transport Layer Security (TLS)~\cite{rfc8446}, an adversary could identify and track users. Since an IPv6 address usually corresponds to one single user rather than a user group due to the rare deployment of NAT, a correlation attack against IPv6 remains a serious individual-level privacy threat. Activity correlation on IPv6 traffic can be grouped into two categories -- address-based correlation and traffic characteristic correlation.

Address-based correlation allows an adversary to associate an IPv6 address with a user's activity, especially when an IPv6 address is in a weak configuration. For example, a user might configure a constant interface identifier through which a user's activity could be pinpointed from multiple contexts~\cite{rfc4862}. To eliminate this address correlation issue, RFC 4291~\cite{rfc4291} standard requires network operators to treat interface identifiers as semantically opaque. RFC 4941 standard~\cite{rfc4941} extends stateless address auto-configuration (SLAAC) allowing IPv6 users to use temporary addresses. 

Different from address-based correlation, traffic characteristic correlation associates traffic with users' activities by analyzing the patterns in the encrypted traffic (e.g.,~\cite{GonzalezSL16, KumpostM09, AndersonM17, NasrBH18}). While demonstrating high effectiveness in traffic-user-correlation, approaches of using this technique can only correlate the traffic of a selected subset of users because of the poor knowledge description and the unreliable similarity learning leads to false positives.

In this paper, we demonstrate that a more sophisticated approach could overcome this limitation, presenting significantly more danger to the privacy of IPv6 users. In particular, we introduce a method to learn a correlation function from TLS encrypted traffic. Using the function, an adversary could determine whether two arbitrary addresses belong to the same user. Unlike prior works, this enables large-scale user activity correlation. Our proposed attack consists of two steps as follows. First, an adversary monitors the TLS encrypted traffic on a vantage point and then constructs a knowledge graph for each client address. Second, using Graph Neural Networks~\cite{WangJSWYCY19} along with Siamese Networks~\cite{ChopraHL05}, we introduce a Siamese Heterogeneous Graph Attention Network -- \textsc{SiamHAN} that employs multi-level attention and metric learning to capture the relationship between two IPv6 addresses with TLS encrypted traffic.

In this work, we evaluate the performance of \textsc{SiamHAN} by using 5-month IPv6 user traffic collected at a vantage observation point. We show that \textsc{SiamHAN} could correlate the activities of pairwise IPv6 addresses with 90\% accuracy based on 1-month adversary's background knowledge. When applied to long-term user tracking and user discovery tasks, \textsc{SiamHAN} outperforms existing correlation techniques by significant margins. For instance, with an adversary's background knowledge on a real-world 5-month dataset, tracking target users or discovering unique users could achieve 99\% or 88\% accuracy. This performance dramatically outperforms the state-of-the-art correlation system Deepcorr~\cite{NasrBH18} which demonstrates only 85\% or 60\% accuracy. 


\textbf{Contributions.} Our contributions can be summarized as:
\begin{itemize}

\item We introduce a new IPv6 address correlation attack that effectively correlates a user's TLS encrypted traffic with its dynamic address.

\item We present a knowledge graph-based approach to model user behavior behind addresses. It exploits multi-type semantic meta-information to facilitate user correlation.

\item We propose a correlation attack model -- \textsc{SiamHAN} which demonstrates superior performance on IPv6 user activity correlation. 

\item We conduct extensive experiments on a 5-month IPv6 user TLS traffic dataset. Results show that \textsc{SiamHAN} is robust and could reach drastically high accuracy on multiple long-term user correlation tasks. 
\end{itemize}

\textbf{Roadmap.} Section \ref{sec2} summarizes the prior researches related to our work. Section \ref{sec3} introduces the threat model and the basic knowledge about IPv6 address correlation attacks. Section \ref{sec4} highlights the overall design of \textsc{SiamHAN}. Section \ref{sec5} presents the main setting for experiments. Section \ref{sec6} shows the evaluation results and Section \ref{sec7} discusses the mitigations against the attack. Section \ref{sec8} concludes the paper.

\section{Related Work}\label{sec2}
From the objective perspective, prior works relevant to ours are user activity correlation. From the technical perspective, the works mostly relevant to ours are heterogeneous graph representation learning as well as metric learning. In the following, we summarize and discuss these works.

\subsection{Address Structure Learning}
One technique to correlate user activities with addresses is to learn address structure and infer user address configuration schemes. RFC 7707 \cite{rfc7707} points out known address configuration schemes and possible administrator configuration customs. The measurement work in the document indicates that most addresses follow specific patterns, which means that even an address configuration with transform addresses, such as DHCPv6 \cite{rfc8415}, may be compromised by address structure learning to narrow the target user range. A body of work \cite{ForemskiPB16,GasserSFLKSHC18,CuiXGSX20} could even learn the addressing pattern through unsupervised clustering or neural networks to facilitate active user discovery. However, RFC 4941 \cite{rfc4941} proposed a temporary address configuration scheme to replace the traditional addressing with pseudo-random interface identifiers that change over time, making it extremely difficult to identify users from the address structure. Ullrich et al. \cite{UllrichW15} analyzed the temporary address generation algorithm and showed that it is possible to infer a user’s future temporary addresses through long-term observations. Because the observation must be conducted on one single user host, address correlation with large-scale users is still impossible to implement. Finally, RFC 7721 \cite{rfc7721} discussed the privacy and security considerations of the IPv6 address generation mechanism and indicated that certain constant information associations might lead to prolonging the observation time of temporary addresses. It is consistent with the idea of our attack. Our work will complete long-term user activity correlation through a combination of address structure learning and traffic characteristics learning.

\subsection{Traffic Characteristics Learning}
Another technique to correlate user activities with addresses is to extract multiple traffic characteristics and then identify users. Since the traffic contains multiple dimensions of user meta-information, moderate learning on the traffic characteristics has proved powerful performance in multiple tasks, including flow correlation \cite{SunEVLRCM15,NasrBH18}, website fingerprinting \cite{SirinamIJW18,Wang20,SirinamMR019}, and traffic classification \cite{LiuHXCL19}. Prior work mainly learns traffic characteristics from three dimensions, including user profiling \cite{GonzalezSL16,KumpostM09}, TLS fingerprinting  \cite{AndersonM17}, and flow sequences \cite{NasrBH18}.

\textbf{User Profiling.} User profiling refers to using behavior-based statistical features to construct a user profile. Kumpost et al. \cite{KumpostM09} employed the target IP address to create user profiles to identify these users in future traffic. Banse et al. \cite{HerrmannBF13} generate user profiles by collecting user DNS requests and utilize a Bayesian classifier \cite{0021593} to track users on the university network. In the case of TLS traffic, Gonzalez et al. \cite{GonzalezSL16} showed that leveraging Server Name Indication (SNI) information could effectively collect user online interests.

\textbf{TLS Fingerprinting.} TLS Fingerprinting is a technique that extracts parameters from a TLS ClientHello to provide visibility into the application that creates the session. Applications of TLS fingerprinting include malware detection \cite{AndersonPM18}, operating system identification \cite{AndersonM17}, and client identification \cite{HusakCJC15}. Due to the difference in fingerprints of different browsers, it is possible to distinguish users to a certain extent. Several open-source databases have been released include \cite{althouse2020ja3,brotherston2020fingerprintls,majkowski2020ssl}.
 
\textbf{Flow Sequences.} Flow sequences are the packet timings and packet sizes collected during the user communications. Nasr et al. \cite{NasrBH18} exploited flow sequences to link the egress and ingress segments of a Tor connection. Liu et al. \cite{LiuHXCL19} could distinguish the application types of user-generated traffic through the characteristic. These works show the effectiveness of flow sequence for user activity identification.

\begin{figure*}[htbp]
\centerline{\includegraphics[width=1.0\textwidth]{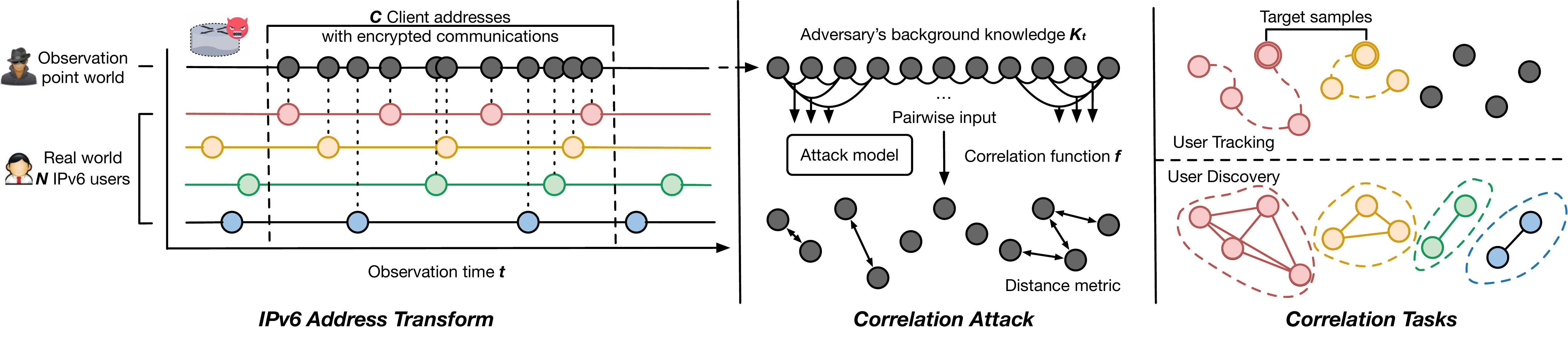}}
\caption{The threat model of IPv6 address correlation attacks. The adversary could collect IPv6 client addresses and their TLS traffic within time $t$ as the background knowledge $\kappa_t$ to conduct correlation learning for user tracking or user discovery.}
\label{fig1}
\end{figure*}

While multi-dimension characteristics could be extracted in TLS encrypted traffic, the complexity of user activity brings extreme difficulty when applied on a large scale. Therefore, prior work is mainly performed on a closed-world dataset, i.e., re-identifying training users on the test set. In this work, we exploit graph-based user profiles to develop the correlation attack model in IPv6 scenarios, achieving strong user correlation on an open-world dataset for unknown users. 

\subsection{Heterogeneous Graph Embedding}
Graph data is a powerful tool to model the complex relationships between entities. Therefore, a family of machine learning techniques known as Graph Neural Networks (GNNs) \cite{KipfW17,HamiltonYL17,VelickovicCCRLB18,WangJSWYCY19} was proposed to analyze graphs. With the growing performance of GNNs and considerable research interest, researchers are no longer satisfied with the study of homogeneous graphs. The real-world graph usually comes with multi-types of nodes and edges, also widely known as Heterogeneous Information Networks (HIN) \cite{ShiLZSY17}. Because the heterogeneous graph contains more comprehensive information and rich semantics, it has been widely used in many data mining tasks \cite{NiuLLXSDC20,WangCYLNTGLCY19}. For instance, Wang et al. \cite{WangJSWYCY19} proposed a Heterogeneous Graph Attention Network (HAN) to provide powerful performance on node classification and clustering tasks. To express the user activities behind the client address in the traffic with sufficient semantic information, we employ a heterogeneous graph as the knowledge of user behavior and exploit a multi-level attention mechanism to learn the graph embedding for correlation tasks.

\subsection{Metric Learning}
Metric learning, also known as distance metric learning, is learning a distance metric for the input space of data from a given collection of pairs of similar or dissimilar entities. The core idea of distance metric is applied by many representative works, including KNN and SVM \cite{WeinbergerBS05}. With the development of deep learning, Siamese Networks \cite{ChopraHL05} is proposed to employ a pair of shared weight network architecture and contrastive loss function to model the distance metric. The simplicity and extensibility of the network structure lead the Siamese Network widely applied in computer vision tasks, including face recognition \cite{TaigmanYRW14} and object tracking \cite{HeLTZ18}. In this paper, we implement a Siamese Network framework based on heterogeneous graph data, called \textsc{SiamHAN}, to acquire a reliable correlation metric between IPv6 client addresses.

\section{Preliminaries}\label{sec3}
This section proposes the threat model of IPv6 address correlation attacks and a brief basic knowledge related to our attack model, including IPv6 addressing and TLS communications, to help readers understand this paper.

\subsection{Threat Model}
Figure \ref{fig1} shows the threat model of IPv6 address correlation attacks. In an IPv6 network, $N$ IPv6 users may generate $C$ client addresses to access online services within a period $t$. There usually is $|N|\leq |C|$ due to frequent changes of the client addresses. However, the relationship between users and addresses cannot be detected using packet contents due to TLS encryption. For instance, $C_i$ and $C_j$ are two IPv6 addresses of a user observed in the traffic during the period $t$. However, such association can not be detected by inspecting the packet contents of $C_i$ and $C_j$ due to TLS encryption.

The goal of an adversary is to correlate two arbitrary IPv6 addresses to identify a unique user. In particular, the adversary could perform an IPv6 address correlation attack, i.e., based on the encrypted communication behavior of all IPv6 addresses for a wiretapping time $t$ as the background knowledge $\kappa_t$, the adversary could judge the relationship $R$ of a pair of addresses $\langle C_i,\, C_j \rangle$ through a correlation function $f$:
\begin{equation}
R = f(\, \langle C_i,\, C_j \rangle \,  |\,  \kappa_t \,)
\end{equation}
The correlation function $f$ can be learned by an attack model providing the distance metric for arbitrary pair of IPv6 addresses, which is used to determine whether they belong to the same user through a threshold $\eta$. 

To train the attack model, the adversary could utilize some tricks to obtain the ground truth dataset, e.g., using leaked plaintext cookies. Although the user data is protected by TLS encryption most of the time, a few users expose their HTTP plaintext since they use the changing addresses and access some websites without HTTPS deployment during adversary's traffic monitoring. Then the adversary could easily label the encryption connections of these addresses through the plaintext cookies. It is worth noting that this situation could only provide a small number of user labels and most users never reveal their plaintext information. Since the attack model only requires training with TLS data, the adversary could perform large-scale correlation attacks on the wild TLS traffic without plaintext once obtaining the model. Another way for the adversary to collect the training set is simulating and generating user data by her own clients. In this case, the adversary might require a more detailed setting to obtain a representative dataset.

An IPv6 address correlation adversary can intercept network traffic at various network locations. According to the different target user groups, these locations could be relay routers, Autonomous systems (ASes), Internet exchange points (IXPs), and website servers.

Since the IPv6 address correlation attack could model an association relationship between any pair of addresses, we consider that the attack could conduct user tracking and user discovery tasks on large-scale encrypted traffic:

\textbf{User Tracking.} Based on the adversary's background knowledge $\kappa_t$, a limited number of target users' one client address activity sample is known. The adversary could correlate all addresses of the target users during the period $t$ to achieve target user tracking, which is like a classification task of classifying all collected samples into target user categories or no correlation categories.

\textbf{User Discovery.} Based on the adversary's background knowledge $\kappa_t$, the number of users in traffic is unknown. The adversary could calculate the correlation between every two addresses and acquire address clusters during the period $t$ to realize user discovery, like a clustering task of classifying all collected samples into unknown user categories.

\subsection{IPv6 Addressing}
An IPv6 address consists of a global routing prefix, a local subnet identifier, and an interface identifier (IID) \cite{rfc4291}. While the global routing prefix is determined to route traffic destined to a Local Area Network (LAN), the configuration of IID is allowed more freedom to ensure the uniqueness of the host interface in the local network segment. RFC 7721 \cite{rfc7721} considers the security and privacy of various address configuration schemes. These schemes show different degrees of privacy threats in the face of address-based correlation:
\begin{itemize}
\item \textbf{Constant IID.} An IPv6 interface identifier that is globally stable, i.e., the IID will remain constant even if the node moves from one IPv6 link to another, which could be generated through IEEE identifier \cite{rfc4862} or static, manual configuration and be used to correlate activities for the device lifetime or the address lifetime.
\item \textbf{Stable IID.} An IPv6 interface identifier that is stable per IPv6 link, i.e., the IID will remain unchanged as long as the node stays on the same IPv6 link but may change when the node moves from one IPv6 link to another, which is described in RFC 7217 \cite{rfc7217} and could be used to correlate activities within single IPv6 link.
\item \textbf{Temporary IID.} An IPv6 interface identifier varies over time. The IID could be generated through SLAAC privacy extension \cite{rfc4941}, or DHCPv6 \cite{rfc8415}, which could only be tracked for the temp address lifetime.
\end{itemize}
Therefore, address-based correlation attacks are usually effective on an address with a constant or stable interface identifier while unachievable on a temporary address. Correlation techniques need more meta-information to overcome dynamic address transform.


\begin{figure*}[htbp]
\centering
\subfigure[Node $C$]{       
\label{1}
\begin{minipage}[t]{0.13\linewidth}
\centering
\includegraphics[width=2.3cm]{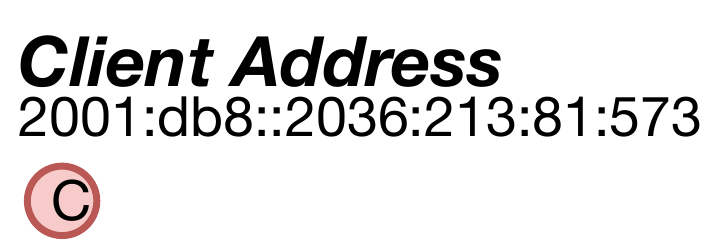}
\end{minipage}
}
\subfigure[Nodes $S$ and $SCS$ meta-paths]{ 
\label{2}
\begin{minipage}[t]{0.26\linewidth}
\centering
\includegraphics[width=4.2cm]{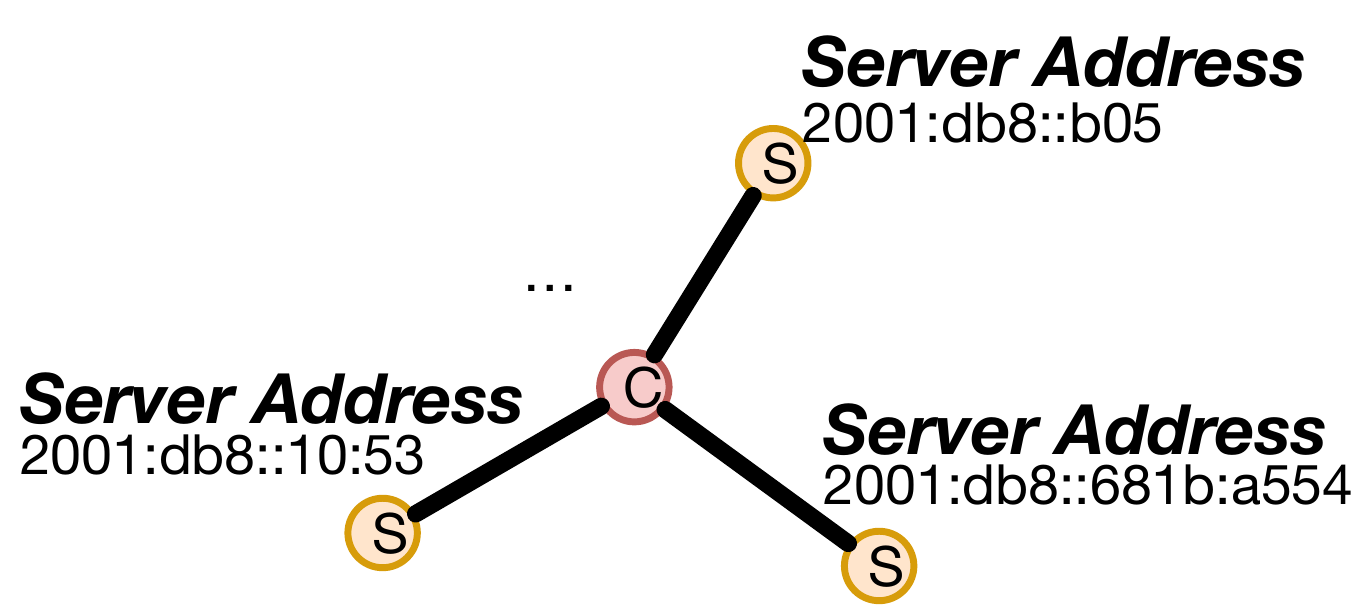}
\end{minipage}
}
\subfigure[Nodes $F$ and $FCF$ meta-paths]{ 
\label{2}
\begin{minipage}[t]{0.26\linewidth}
\centering
\includegraphics[width=4.7cm]{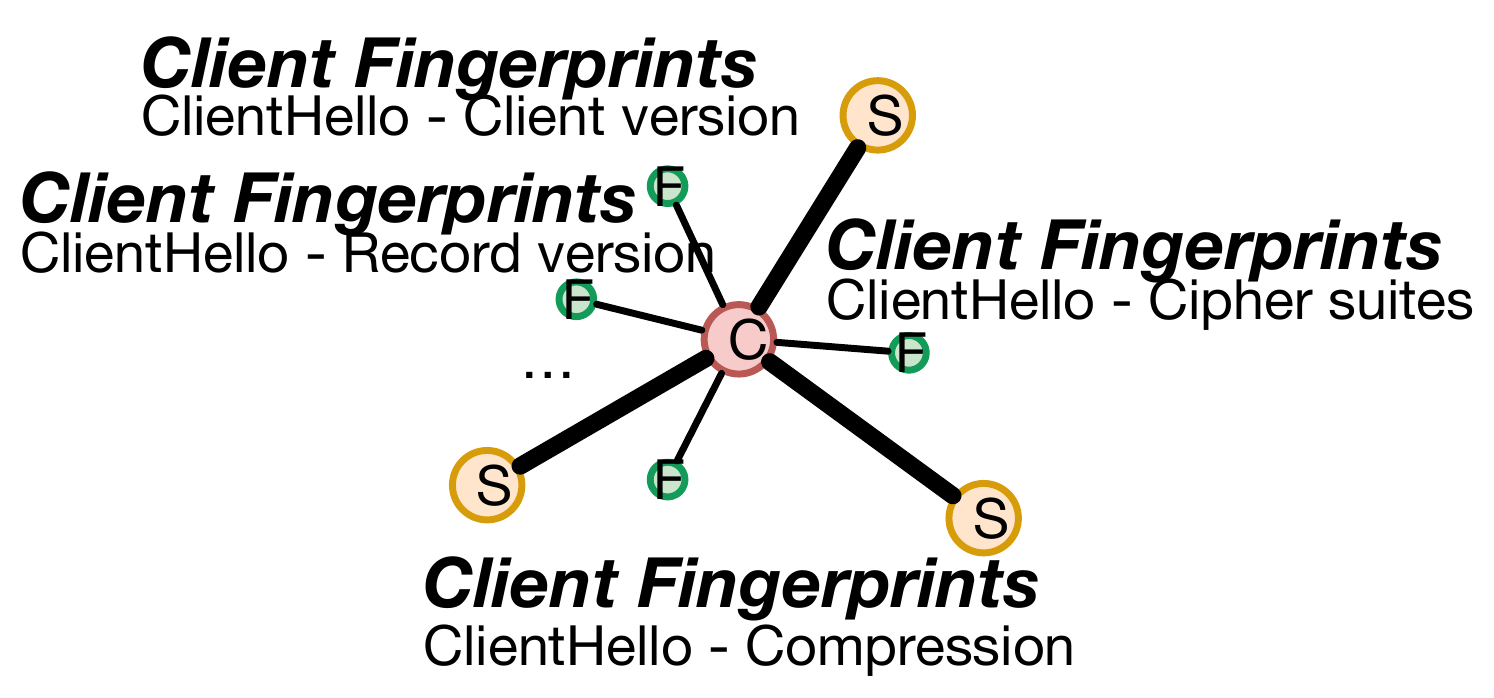}
\end{minipage}
}
\subfigure[Nodes $F$ and $FSF$ meta-paths]{ 
\label{2}
\begin{minipage}[t]{0.26\linewidth}
\centering
\includegraphics[width=5.7cm]{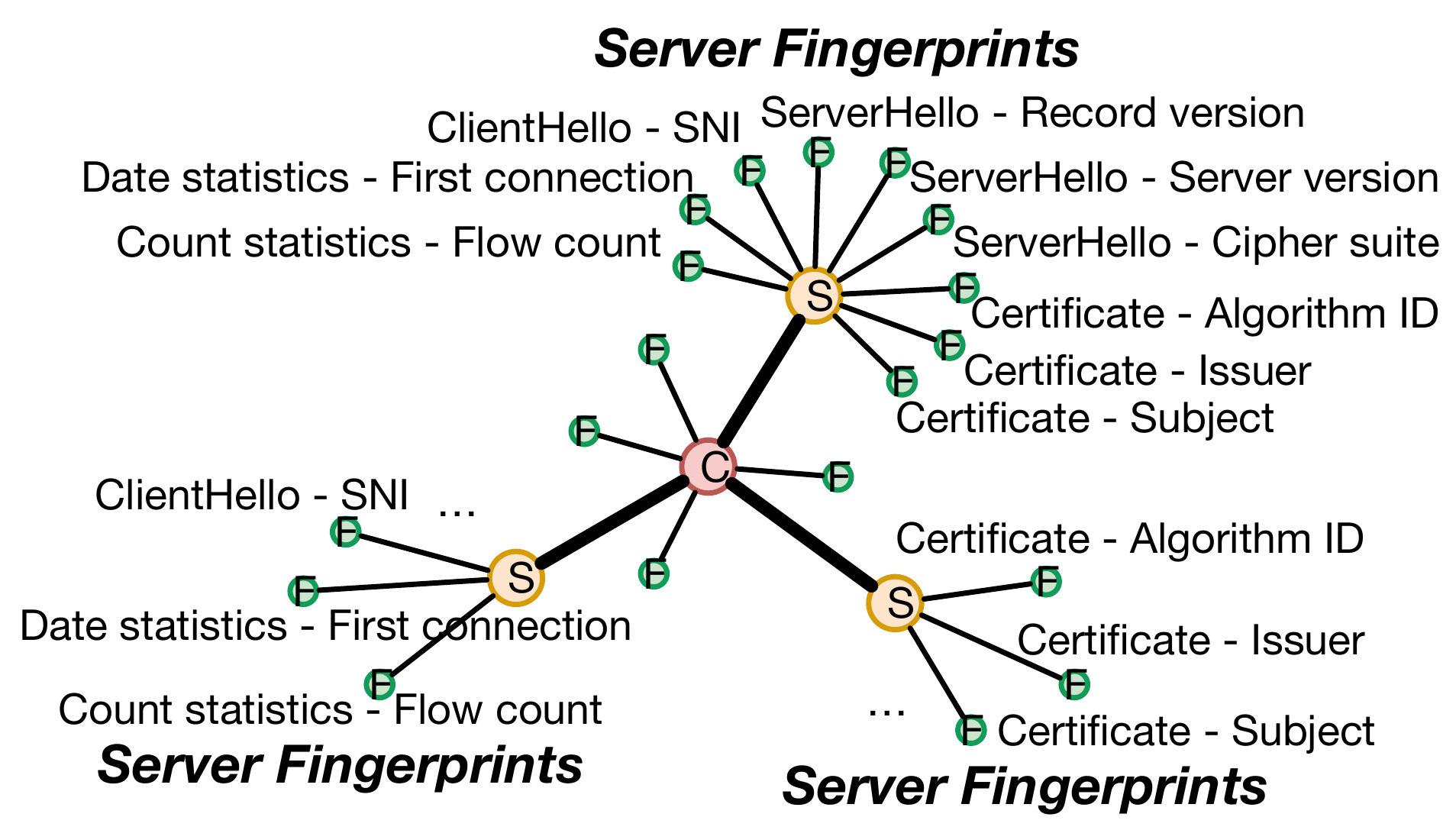}
\end{minipage}
}
\centering
\caption{The building process of the knowledge graph on one IPv6 client address.}
\label{fig2}
\end{figure*}

\begin{table}[tb]
\caption{The notions of the TLS fields used in the paper.}
\begin{center}
\begin{tabular}{rl}
\toprule
\textbf{Field Name}  & \textbf{Notion}\\
\midrule
Record version&The version of the TLS protocol\\
& employed in the Record Layer.\\
Client version&The version of the TLS protocol by\\
&which the client wishes to communicate.\\
Server version&The version of the TLS protocol \\
&finally chosen by the server.\\
Cipher suites&A list of the cryptographic options \\
& supported in the ClientHello or the single \\
& cipher suite selected in the Server Hello.\\
Compression&A list of the compression methods\\
& supported in the ClientHello or the single\\
& method selected in the ServerHello.\\
SNI& The domain name specified by the client\\
& to reach in the ClientHello extensions.\\
Algorithm ID& The identifier for the cryptographic\\
& algorithm used to sign the certificate.\\
Issuer& The entity that has signed and issued the\\
&  certificate in the Certificate message.\\
Subject&The entity associated with the public \\
& key stored in the Certificate message.\\
\bottomrule
\end{tabular}
\label{tab1}
\end{center}
\end{table}

\subsection{TLS Communication}
TLS \cite{rfc8446} is an encryption protocol designed to secure Internet communications. Whenever a user navigates to a website over HTTPS, the TLS encryption of the traffic payload effectively protects user privacy from malicious analysis. However, before the encrypted communications, a TLS handshake is required to exchange several messages to establish the session of the two communicating sides, which includes considerable available meta-information to infer activities. For instance:
\begin{itemize}
\item \textbf{ClientHello message.} The client initiates the handshake by sending a ClientHello message to the server. The message includes which TLS version the client supports, the cipher suites supported, the compression methods supported, a random string, and the extension field. Clients may request extended functionality from servers by sending data in the extensions field, like specifying Server Name Identifier (SNI) to prevent common name mismatch errors.
\item \textbf{ServerHello message.} In reply to the ClientHello message, the server sends a ServerHello message containing a random string, extensions, the server's chosen TLS version, cipher suite, and compression method.
\item \textbf{Certificate message.} The Certificate message will always immediately follow the ServerHello message when required certificates for authentication, which conveys the server's certificate chain to the client. The certificate also contains the meta-information related to the server, such as issuer and subject.
\end{itemize}
To help readers understand this paper, we provide the notions of the TLS fields related to the paper in Table \ref{tab1}. Based on the meta-information proposed from the TLS communications, a user's communication activities could be learned due to the exposure of client's and server's information in the session. However, complex user activities and considerable field information render correlation attacks infeasible in multiple contexts. Advanced traffic characteristic correlation techniques must focus on favorable information to facilitate effective user correlation.

\section{Design of \textmd{\textsc{SiamHAN}}}\label{sec4}
This section introduces our IPv6 address correlation attack system, called \textsc{SiamHAN}, which is a two-step attack, including building knowledge graphs and learning attack models.


\begin{table}[tb]
\caption{The detail of nodes in the knowledge graph.}
\begin{center}
\begin{tabular}{llll}
\toprule
\textbf{Node Type} & \textbf{Source} & \textbf{Label} & \textbf{Node Attribute}\\
\midrule
Client node& IPv6 header & $C$ & Client address\\
\hline
Server node & IPv6 header & $S$& Server address\\
\hline
 & \multirow{4}{*}{ClientHello} & $F_1$&Record version\\
Client  && $F_2$&Client version\\
fingerprint  && $F_3$&Cipher suites\\
&& $F_4$&Compression\\
\hline
&ClientHello& $F_5$ &SNI\\
\cline{2-4}
&&$F_6$&Record version\\
&ServerHello&$F_7$&Server version\\
Server&&$F_8$&Cipher suite\\
\cline{2-4}
fingerprint &&$F_9$& Algorithm ID\\
&Certificate&$F_{10}$&Issuer\\
&&$F_{11}$&Subject\\
\cline{2-4}
&Date statistics&$F_{12}$&First connection\\
&Count statistics&$F_{13}$&Flow count\\
\bottomrule
\end{tabular}
\label{tab2}
\end{center}
\end{table}

\subsection{Knowledge Graph}
When chronically intercepting network traffic on the victim router or server, the adversary could collect considerable meta-information about the client address communication, which could be reconstructed to help identify the user. To achieve this goal on the IPv6 network, we construct a knowledge graph based on TLS encrypted communication for each IPv6 client address as the adversary's background knowledge $\kappa_t$. Since the user’s complex online behavior will generate diverse semantic data during the adversary’s wiretapping time $t$, we use a heterogeneous graph \cite{SunH12} to model the knowledge graph. It contains multi-type nodes and neighbor relationships to describe the user activities behind the address accurately. Figure \ref{fig2} shows the building process of the knowledge graph.

\textbf{Node and Node Attribute.} Based on the adversary's background knowledge $\kappa_t$, the knowledge graph of each IPv6 client address contains three types of nodes, including a client node $C$, server nodes $S$, and fingerprint nodes $F$, which are shown in Table \ref{tab2}. Each graph node keeps an attribute to represent the meaning of the node:

\begin{itemize}
\item \textbf{Client node $C$.} The client node models an IPv6 client address that is monitored within time $t$, whose attribute is the 32-digit hexadecimal IPv6 client address. Each knowledge graph contains only one client node to denote the meta-information related to it.
\item \textbf{Server node $S$.} The server nodes are all IPv6 server addresses that have established TLS communications with the client address, whose attribute is the 32-digit hexadecimal IPv6 server address.
\item \textbf{Fingerprint node $F$.} The fingerprint nodes include client fingerprints and server fingerprints, whose attributes are field values of the ClientHello, ServerHello, Certificate messages, and statistical characteristics in the TLS connection established with the client address. Following the work of \cite{althouse2020ja3,brotherston2020fingerprintls,majkowski2020ssl}, we intend to select the commonly used, distinguishable TLS fields for model learning. In addition, the statistical characteristics provide a more detailed description of the user behavior. \texttt{First connection} refers to the date of the first time the client accesses the server. \texttt{Flow count} records the number of the flow generated during the communication.
\end{itemize}
Because the attributes of these nodes integrate address and traffic characteristic meta-information, the adversary could learn user activities based on the knowledge of address structure and traffic characteristic correlation.

\textbf{Neighbor Relationship.} In a heterogeneous graph, nodes can be connected via different semantic paths, which are called meta-paths \cite{SunHYYW11}. To denote the neighbor relationship of different semantics, we propose three types of meta-paths to connect three types of nodes in the knowledge graph, including $SCS$ meta-path, $FCF$ meta-path, and $FSF$ meta-path:
\begin{itemize}
\item \textbf{$SCS$ meta-path.} The $SCS$ meta-path connects the client node $C$ and multiple server nodes $S$, which represents the TLS communication activities between the client and multiple servers.
\item \textbf{$FCF$ meta-path.} The $FCF$ meta-path connects the client node $C$ and multiple client fingerprint nodes $F$, which represents the browser parameters that may be used behind the client.
\item \textbf{$FSF$ meta-path.} The $FSF$ meta-path connects each server node $S$ and multiple server fingerprint nodes $F$ related to the server, which denotes the service characteristics behind each server. 
\end{itemize}

The $FCF$ meta-path and the $FSF$ meta-path can be effectively exploited to learn unique client and service representations. The $SCS$ meta-path exposes the communication activities between the user's client and each service, thus facilitating correlation attacks reliable.

It is worth noting that, since the user may use multiple browsers, the same type of client fingerprint may contain multiple nodes, e.g., two client fingerprint nodes with different Cipher suites attributes are included in one knowledge graph. In addition, since a TLS connection may not contain all three types of messages, a server node could lack some server fingerprints, thus leaving a smaller count of the $FSF$-based neighbors.

\begin{figure*}[htbp]
\centerline{\includegraphics[width=1.0\textwidth]{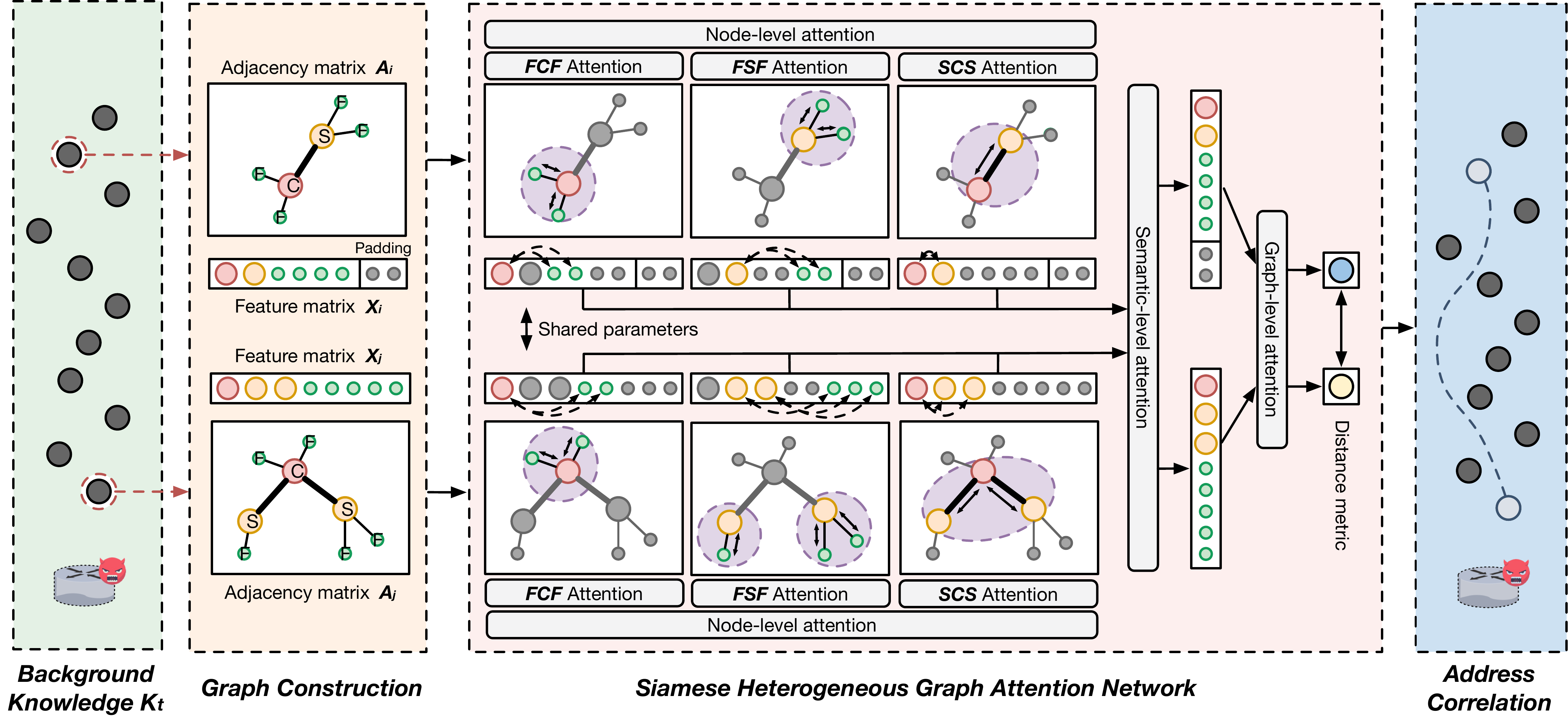}}
\caption{The overall architecture of \textsc{SiamHAN}. \textsc{SiamHAN} inputs pairwise client addresses' knowledge graphs to learn their correlation. The multi-level attention mechanism helps match the similar features between the two heterogeneous graphs to learn their graph embeddings. The Siamese Network finally metric the distance of the embeddings to judge the correlation relationship.}
\label{fig3}
\end{figure*}

\subsection{Model Architecture}
\textsc{SiamHAN} is a deep learning framework shown in Figure~\ref{fig3}, which exploits the Heterogeneous Graph Attention Network-based Siamese Network architecture to learn the address correlation. The architecture could be divided into four objectives: node-level attention, semantic-level attention, graph-level attention, and metric learning with Siamese Network.

After constructing a knowledge graph for each client address, the adversary could select any two knowledge graphs to model their association and infer whether they are bound to the same user. Each knowledge graph could extract an adjacency matrix $A$ and feature matrix $X$ to be processed by the GNNs, where the adjacency matrix $A$ includes the neighbor relationships of each node and the feature matrix $X$ is the attribute value of all nodes. A GNN method iteratively updates a node’s features via aggregating its neighbors’ features. \textsc{SiamHAN} uses self-attention \cite{VaswaniSPUJGKP17} with multiple levels to update a pair of feature matrix $X_i$ and $X_j$ according to the adjacency matrix $A_i$ and $A_j$ and obtain their network embeddings to measure the distance of the two addresses for correlation.

\textbf{Node-level Attention.} For each pair of input knowledge graphs, the node-level attention first learns the weights of meta-path-based neighbors and aggregates them to get the semantic-specific node embedding. Given $N_u^\Phi$ denotes the meta-path $\Phi$ based neighbors of node $u$ (include itself) and node $v \in N_u^\Phi$, the importance of meta-path-based node pair $\langle u,  v \rangle$ can be formulated as follows:
\begin{equation}
\begin{gathered}
e_{uv}^\Phi = \sigma (a_\Phi^T \cdot [h_u || h_v]),\\
\alpha_{uv}^\Phi = {\rm softmax_v}(e_{uv}^\Phi) = \frac{{\rm exp}(e_{uv}^\Phi)}{\sum_{k\in N_u^\Phi}{\rm exp}(e_{uk}^\Phi)},
\end{gathered}
\end{equation}
where $h_u$ and $h_v$ are the features of node $u$ and $v$, $a_\Phi$ is the node-level attention parametrize matrix for meta-path $\Phi$, $\sigma$ denotes the activation function, and || denotes the concatenate operation. Since the attack model inputs pairwise user meta-information, a larger weight coefficient $\alpha_{uv}^\Phi$ indicates matching similar neighbor data in a single meta-path-based semantic in the two knowledge graph, which contributes to the correlation task. For instance, the two client nodes in the pairwise graph link to the same server nodes based on $\Phi_{SCS}$.


Then, the meta-path-based embedding of node $u$ can be obtained by aggregating all neighbor attributes with the corresponding coefficients as follows:
\begin{equation}
z_u^\Phi = \Big \|_{k=1}^{K}\sigma \Big(\sum_{v\in N_u^\Phi}\alpha_{uv}^\Phi\cdot h_v \Big),
\end{equation}
where $z_u^\Phi$ is the learned embedding of node $u$ for the meta-path $\Phi$, $K$ is the head number using the multi-head attention mechanism \cite{VaswaniSPUJGKP17}. Among the three types of meta-path in our work, $FCF$ and $FSF$ promote learning the unique client and server service embeddings based on the client and server fingerprints, while $SCS$ mines the user activity representation using the communication relationship.

\textbf{Semantic-level Attention.} After feeding node features into node-level attention with the meta-path set $M=\{\Phi_{FCF}, \Phi_{FSF}, \Phi_{SCS}\}$, the semantic-level attention is required to learn the importance of three types of semantic-specific embeddings and fuse them as a comprehensive node embedding. The importance of meta-path $\Phi_i$ based embedding is shown as follows:
\begin{equation}
\begin{gathered}
w_{\Phi_i} = \frac{1}{|V|}\sum_{u\in V}p^T\cdot {\rm tanh}(W_s \cdot z_u^{\Phi_i} + b_s),\\
\beta_{\Phi_i} = {\rm softmax_i}(w_{\Phi_i}) = \frac{{\rm exp}(w_{\Phi_i})}{\sum_{\Phi_i \in M}{\rm exp}(w_{\Phi_i})},
\end{gathered}
\end{equation}
where $W_s$ is the weight matrix, $b_s$ is the bias vector, and $p$ is the semantic-level attention parametrize matrix. $V$ is the node-set of the input knowledge graph. Since we average the importance of all the semantic-specific node embedding, the weight coefficient $\beta_{\Phi_i}$ could be interpreted as the contribution of the meta-path $\Phi_i$ for the correlation task.

With the learned weights of each semantic-specific embedding, the comprehensive embedding $s_u$ of node $u$ could be:
\begin{equation}
s_u = \sum_{\Phi_i \in M}\beta_{\Phi_i} \cdot z_u^{\Phi_i}.
\end{equation}
The comprehensive embeddings are the final representations of nodes learned by \textsc{SiamHAN}, which aggregates multiple semantic characteristics. For instance, the client node $s_C$ finally obtains the embedding with the semantics that using specific browsers to access online services by integrating semantic-specific embeddings $z_C^{\Phi_{FCF}}$ and $z_C^{\Phi_{SCS}}$.

\textbf{Graph-level Attention.} To gain the graph embedding for distance metric learning, graph-level attention is proposed to aggregate the final embeddings of all nodes in the knowledge graph.  The importance of node $u$ could be obtained as follow:
\begin{equation}
\begin{gathered}
g_u = q^T \cdot {\rm tanh}(W_g \cdot s_u + b_g),\\
\gamma_u = {\rm softmax_u}(g_u) = \frac{{\rm exp}(g_u)}{\sum_{u \in V}(g_u)},
\end{gathered}
\end{equation}
where $W_g$ is the weight matrix, $b_g$ is the bias vector, and $q$ is the graph-level attention parametrize matrix. Unlike $\alpha_{uv}^\Phi$ with local attention on neighbors, a larger weight coefficient $\gamma_u$ denotes globally matching similar nodes in the two knowledge graphs. Therefore, the graph embedding $Z$ could be formulated as follows:
\begin{equation}
Z = \sum_{u \in V}\gamma_u \cdot s_u
\end{equation}

\textbf{Metric Learning with Siamese Network.} The goal of the Siamese Network architecture in our work is to metric the distance $D$  between the knowledge graph of two arbitrary IPv6 client addresses, which could be used to judge the correlation relationship $R$ through a threshold $\eta$:
\begin{equation}
\begin{gathered}
D = ||Z_1 - Z_2 ||_2,\\
R = \left\{ 
\begin{aligned}
1 &&D<\eta\\
0 &&D\geq\eta\\
\end{aligned}
\right.,
\end{gathered}
\end{equation}
where $Z_1$ and $Z_2$ are the final graph embeddings of the two input knowledge graphs. $R=1$ means the two client addresses come from the same user, otherwise $R=0$.

To train the attack model for IPv6 address correlation, \textsc{SiamHAN} requires sets of positive samples and negative samples to learn the correlation function. The positive samples are the pairwise knowledge graphs of IPv6 client addresses bound to the same user. The negative samples are arbitrary pairs of knowledge graphs that come from two different IPv6 users. With both negative and positive samples in hand,  \textsc{SiamHAN} could then be optimized by minimizing the contrastive loss $L$:
\begin{equation}
L = Y\cdot D^2 + (1-Y)\{{\rm max}(0, m-D)\}^2,
\end{equation}
where $Y$ are the labels of the input samples, margin $m$ is a hyperparameter to control the maximum distance that can be considered to update the network. Furthermore, the network parameters are shared in the pairwise network architecture, which focuses on the input difference and learns the similarity. 

\section{Experiment Setup and Implementation}\label{sec5}
This section discusses our dataset collection and composition, compared baselines, evaluation metrics, and the model implementation.

\subsection{Data Collection} 
We passively monitored the IPv6 user traffic from March to July 2018 on China Science and Technology Network (CSTNET) and collect an extensive user dataset for attack experiments. We intend to utilize persistent HTTP plaintext cookies to label the TLS traffic of the addresses that leak these cookies. This situation comes from that these users accessed some websites with HTTPS deployment and used the same address to access some other websites without HTTPS deployment during our traffic monitoring. Therefore, we mainly collected 5-month HTTP and TLS traffic to build the ground truth dataset. Firstly, we searched persistent cookies used continuously during the monitoring period through the Cookie field in the HTTP plaintext to label the frequently transforming client addresses, which obtains considerable address lists corresponding to unique cookies. Note that we only record the persistent cookies that are generated on the first day and continuously used during the observation. We do not consider new cookies that appear during the monitoring period to prevent biases from the accuracy of user labeling. Secondly, since a user usually generates multiple cookies, we aggregated the lists with the same addresses to acquire unique IPv6 HTTP users and their client addresses. Finally, we searched for the communication records of these addresses in the TLS traffic, thus obtaining encrypted communication data of 1.7k users and 2.6k addresses generated by them. Since \textsc{SiamHAN} requires pairwise client address knowledge graphs as input, we combine any two addresses into pairs in the training, validation, or test set. In addition, we generate the correlation labels of the pair samples according to their user labels, which results in 1.5M pair samples. The correlation label is 1 when the two addresses belong to the same user. Otherwise, the correlation label is 0. 

\begin{table*}[htb]
\caption{The analysis of the 5-month real-world user dataset with 4 dimensions including the top ASes of client addresses, the top OSes of user devices, the top SNI accessed by users, and the prevalence of the TLS fields used in the paper. The prevalence of Record version and Cipher suites are shown as the percent of the field in ClientHello/ ServerHello.}
\begin{center}
\begin{tabular}{ll|ll|ll|ll}
\toprule
\textbf{AS Name}  & \textbf{\%Hits} & \textbf{Device OS} & \textbf{\%Hits} & \textbf{SNI} & \textbf{\%Hits} & \textbf{TLS Field} & \textbf{\%Hits}\\
\midrule
CSTNET&78.6\%&Windows&63.7\%&*.google.com&17.9\%&Record version&93.1\%/ 93.9\%\\
China Unicom&10.1\%&Android&23.7\%&*.adobe.com&11.6\%&Client version&93.1\%\\
CNGI-CERNET2&4.0\%&iOS&6.2\%&*.microsoft.com&11.2\%&Server version&93.9\%\\
CERNET&2.4\%&Linux&5.0\%&*.gstatic.com&4.8\%&Cipher suites&93.1\%/ 93.9\%\\
Reliance Jio&1.6\%&Mac OS X&1.3\%&*.macromedia.com&3.3\%&Compression&93.1\%\\
Cloudflare&0.8\%&BlackBerry&0.1\%&*.cloudflare.com&2.4\%&SNI&93.1\%\\
PKU6-CERNET2&0.5\%&Chrome OS&0.1\%&*.2mdn.net&1.9\%&Algorithm ID&78.4\%\\
TSINGHUA6&0.5\%&Symbian OS&0.1\%&*.xboxlive.com&1.6\%&Issuer&78.4\%\\
ZSU6-CERNET&0.4\%&Firefox OS&0.1\%&*.xhcdn.com&1.2\%&Subject&78.4\%\\
\bottomrule
\end{tabular}
\label{tab3}
\end{center}
\end{table*}

\begin{table*}[htb]
\caption{The average number of meta-information per knowledge graph with 1-month background knowledge and the statistics of the dataset with a time-based split evaluated in most experiments of the paper.}
\begin{center}
\begin{tabular}{llll|llll}
\toprule
\textbf{Meta-Path $\Phi$} & \textbf{Relations (A-B)}  & \textbf{Number of A} & \textbf{Number of B} & \textbf{Entity} & \textbf{Training} & \textbf{Validation} & \textbf{Test}\\
\midrule
$SCS$&Client-Server&1.0&5.4&User&1.0k&0.2k&0.5k\\
$FCF$&Client-Fingerprints&1.0&3.8&Sample Pair&1.2M&0.1M&0.2M\\
$FSF$&Server-Fingerprints&5.4&41.3&Knowledge&3 months&1 month&1 month\\
\bottomrule
\end{tabular}
\label{tab4}
\end{center}
\end{table*}

\subsection{Collection Ethics}
To protect user privacy from being exposed in our attack experiments, we anonymize all addresses collected in our dataset with 2001:db8::/32 documentation reserved prefix according to RFC 3849 \cite{rfc3849}. All the plaintext cookies for user labeling are encrypted due to our anonymization work. We did not over-explore data involving personal information based on the Internet measurement standard during the traffic monitoring period. We design a detailed exit mechanism to remove user traffic and record only necessary traffic characteristics used in our experiments. Therefore, our attack experiments are performed to correlate the users with virtualized user IDs rather than linking to real-world individuals. We further guarantee that our measurements do not disrupt or harm evaluation targets. Our work has been approved by our institutional ethics review body to ensure ethical soundness and justification. 

\subsection{Dataset}
We evaluate \textsc{SiamHAN} on the real-world dataset collected through 5 months of traffic monitoring. To indicate the practicality of the attack, we provide a deep eye on the dataset composition and discuss the feasibility of the experiment. 

\textbf{Basic Composition.} After a long-time of data collection, we analyze the basic dataset composition in Table \ref{tab3}. Results show that the dataset keeps a good variation to evaluate the attack model. \textbf{(1) User Source.} We first analyze the top ASes and the percent of addresses within each. During the observation, in addition to the main source of CSTNET, the collected users come from diverse IPv6 networks, including mobile networks (e.g., China Unicom), CDN networks (e.g., Cloudflare), and university networks (e.g., PKU6-CERNET2). \textbf{(2) Client Device.} The users are labeled by the HTTP plaintext in our dataset. Thus, we additionally analyze their HTTP user-agent to infer their device OSes. Results indicate that IPv6 users are using multiple types of devices during the monitoring. \textbf{(3) Online Habit.} Considering the top SNI accessed by users, although the top three domains have the largest number of visits, the access rate of each domain is not high, which indicates that IPv6 users maintain a wide range of online habits and complex behaviors in our dataset. 

\textbf{Graph Samples.}  \textbf{(1) Meta-information.} To better understand the composition of the meta-information in a knowledge graph, we show the statistics of the knowledge graph built for each client address with 1-month background knowledge in Table \ref{tab4}. Since most client addresses are short-lived due to IPv6 address transform, each client address is used to access an average of 5 $\sim$ 6 online services during the 1-month observation. In addition, considering the observation bias caused by issues like packet loss, a few TLS connections do not contain ClientHello, ServerHello, or Certificate message, which lead to an average of 3.8 client fingerprint nodes and 41.3 server fingerprint nodes in each knowledge graph. \textbf{(2) Time-based Data Split.} To simulate a realistic setting implemented by an adversary, we evaluate \textsc{SiamHAN} with a time-based split on the 5-month dataset in most experiments of the paper (except timeliness evaluation in Section \ref{sec6.3}), which uses the first 3-month data for training, the 4th month's data for validation, and the 5th month's data for test. Note that the test user is excluded from the training set. The adversary's intention is to train the attack model on the history ground truth dataset and perform the attack on the future collected data with the background knowledge.

\textbf{Feasibility Discussion.} There may remain doubts about the dataset that could be discussed: \textbf{(1) Labeling Trick.} Since the leaked plaintext cookies are available to label the TLS users in our dataset, one of the doubts could be the necessity of the adversary to train \textsc{SiamHAN}. According to the statistics, we collect a total of 0.58M addresses in the 5-month TLS traffic. Users of only 2.9k addresses expose their HTTP cookies (including the new generated cookies during the observation), which is a ratio of 0.5\%. Therefore, most TLS users never reveal their plaintext cookies, which shows a strong motivation to perform the attack. \textbf{(2) User Assumption.} Our dataset is all composed of users with plaintext cookies. Therefore, another doubt is whether the assumption that users without any plaintext cookies would behave the same as users with plaintext cookies is reasonable. To address the issue, we additionally provide an analysis of users without plaintext cookies in Appendix \ref{app1}. The similar user source and online habits indicate that the model trained on the users with plaintext cookies could be directly generalized to perform effective attacks on the TLS users without plaintext cookies. \textbf{(3) Feature Prevalence.} In Table \ref{tab3}, we analyze the prevalence of the TLS fields used in the paper. The frequent appearance of most fields in the TLS connections ensures that the fingerprints are sufficient to perform the attack. Although more knowledge graphs may lack the fingerprints related to the Certificate message, the knowledge volume composed of other features is still enough to perform the strong correlation attack.

\subsection{Baselines} 
The prior work mainly performs user correlation under TLS traffic from the dimensions including user profiling \cite{GonzalezSL16,KumpostM09}, TLS fingerprinting \cite{AndersonM17}, and flow sequences \cite{NasrBH18}. Among them, we implement four representative methods to compare with \textsc{SiamHAN}: \textbf{(1) User IP Profiling} \cite{GonzalezSL16}\textbf{.} User IP Profiling is building user profiles through all the destination IPs of the client address and using a Bayesian classifier \cite{0021593} to identify known users in a closed-world dataset. To apply address correlation to identify unknown users in an open-world scenario, we use pairwise profiles as input to the classifier to evaluate the performance of correlation attacks. \textbf{(2) User SNI Profiling} \cite{KumpostM09}\textbf{.} User SNI Profiling is to use the SNIs in all the TLS ClientHello messages from the client as a user interest identification. Similar to Banse et al. \cite{HerrmannBF13}, we also exploit a Bayesian classifier to input pairwise SNI profiles to correlate user activities. \textbf{(3) Client Fingerprinting} \cite{AndersonM17}\textbf{.} Client Fingerprinting is to extract the specific fields of the TLS ClientHello message as the user's client fingerprints and leverages Random Forest \cite{Breiman01} to learn the correlation of any two paired fingerprints. \textbf{(4) Deepcorr} \cite{NasrBH18}\textbf{.} Deepcorr uses the flow sequence characteristics to achieve correlation tasks in multiple scenarios \cite{WangR03,NasrBH18,BahramaliHSGT20}. To keep the same setup with Deepcorr, we also extract a flow sequence of 300 packets per client address to indicate the performance of Deepcorr.


\subsection{Evaluation Metrics}
Our evaluation metrics include true positive rate, false-positive rate, area under ROC curve, and accuracy: \textbf{(1) True Positive Rate (TPR).} The TPR measures the fraction of associated address pairs that are correctly declared to be correlated by \textsc{SiamHAN}. \textbf{(2) False Positive Rate (FPR).} The FPR measures the fraction of non-associated address pairs that are mistakenly identified as correlated by \textsc{SiamHAN}. \textbf{(3) Area Under Curve (AUC).} The AUC metric is to calculate the area under the ROC curve formed by TPRs and FPRs with multiple thresholds, which is frequently used in binary classification tasks. \textbf{(4) Accuracy.} To evaluate the user tracking and user discovery tasks applied by correlation attacks, we define two task-based accuracy metrics to indicate the attack performance. In the user tracking task, the \textbf{Tracking Accuracy (TA)} is used to measure the fraction of correctly identified address pairs associated or non-associated with the target user samples. While in the user discovery task, the \textbf{Discovery Accuracy (DA)} is used to evaluate the fraction of addresses that are correctly classified into unique user groups.


\subsection{Implementation}
During the data preprocessing, we limit the maximum node number to 50 and the maximum node attribute length to 50 in each knowledge graph. As such, the dimension of the adjacency matrix $A$ and the feature matrix $X$ is 50$\times$50. The matrix contains padding or truncating operations due to the difference between the number of each graph's nodes or the attribute length of each node. The character in node attributes is encoded to digital and the feature matrix $X$ finally requires row-normalization as the input of  \textsc{SiamHAN}. When training \textsc{SiamHAN}, we randomly initialize parameters and optimize the model with Adam \cite{KingmaB14}, we set the learning rate to 0.005, the regularization parameter to 0.001, and the dimension of the hierarchical attention parametrize matrix $a$, $p$, $q$ to 100, 128, 128. In addition, we also set the number of attention head K to 4. The activation function $\sigma$ is LeakyReLU. The margin $m$ is 20, and the threshold $\eta$ for address correlation is 10. We use early stopping with a patience of 100 to train the model.

\section{Evaluation}\label{sec6}
This section presents the analysis of IPv6 address transform scenarios and all experimental results showing the effectiveness of correlation attacks conducted by \textsc{SiamHAN}.


\subsection{Analysis of Address Transform}
\begin{figure}[tbp]
\centerline{\includegraphics[width=0.5\textwidth]{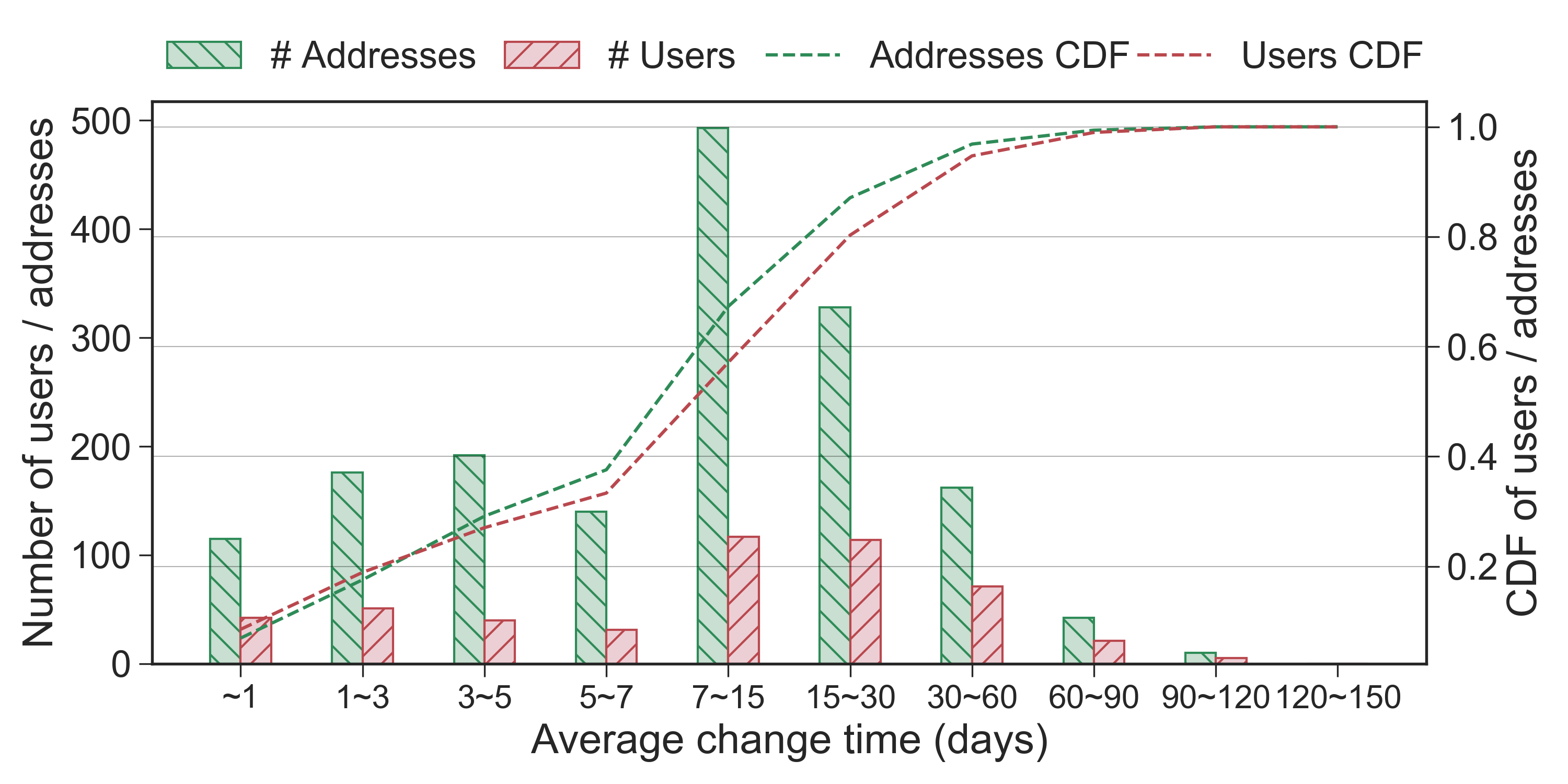}}
\caption{The number of users and their addresses with different average change times under the 5-month TLS traffic.}
\label{fig4}
\end{figure}

During the 5-month traffic monitoring, IPv6 users frequently updated their client addresses for communication. Figure \ref{fig4} reveals the average transform time of these client addresses. 80\% of IPv6 users change their client addresses at least once a month, showing the universality of address transform in IPv6 networks. Considerable users frequently change the client address within two weeks. Since RFC 4941 \cite{rfc4941} recommends the lifetime of the temporary address is one day to one week, the appearance in the TLS traffic is consistent with the instruction. Furthermore, we also analyze the addressing scheme of the address dataset. 98\% of changing user's addresses update the subnet identifier, which explains the mobility of users because the Regional Internet Registry (RIR) usually sets this identifier for different regions. In addition, 23\% of user's addresses maintain a constant IID, while 77\% of IPv6 users use a stable or temporary IID. The analysis results demonstrate that the transform of IPv6 client addresses leads to difficulty in conducting effective user correlation under TLS traffic. 

\subsection{Analysis of Hierarchical Attention} 
\begin{figure}[tb]
\centering
\subfigure[$SCS$ attention of node $C$]{ 
\begin{minipage}[t]{0.47\linewidth}
\centering
\includegraphics[width=4.25cm]{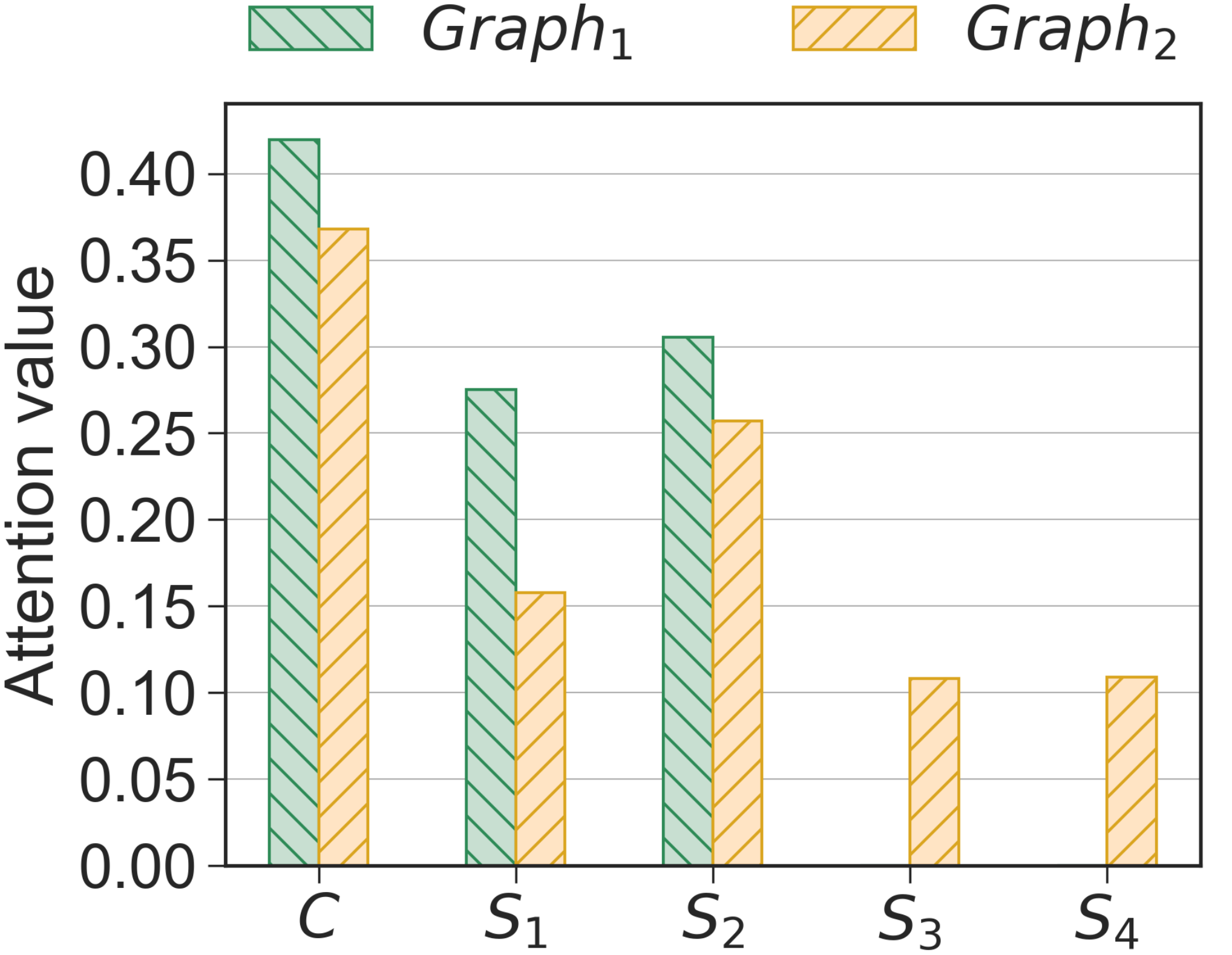}
\end{minipage}
\label{5a}
}
\subfigure[$FCF$ attention of node $C$]{ 
\begin{minipage}[t]{0.47\linewidth}
\centering
\includegraphics[width=4.25cm]{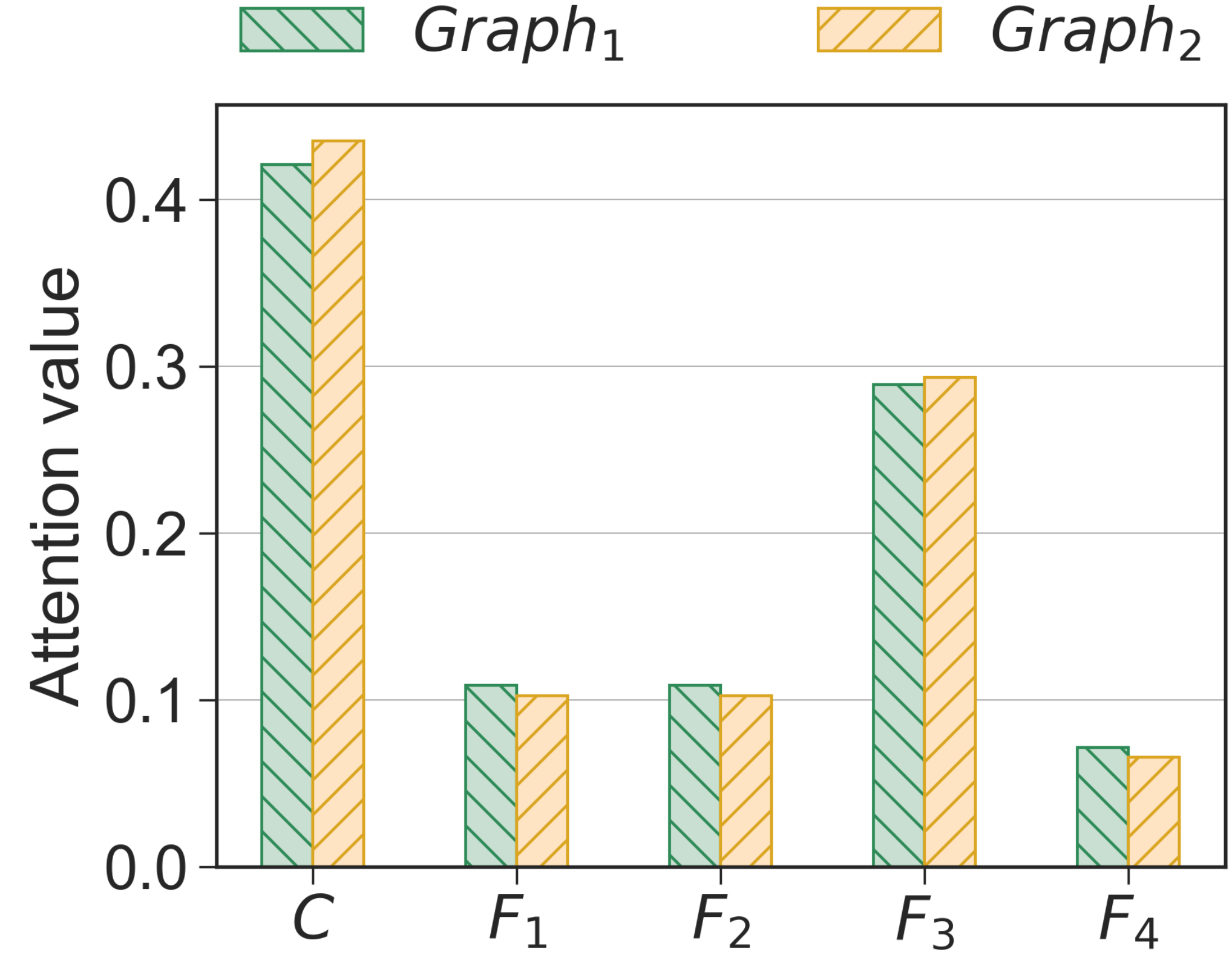}
\end{minipage}
\label{5b}
}
\centering
\subfigure[$FSF$ attention of node $S_2$]{ 
\begin{minipage}[t]{0.63\linewidth}
\centering
\includegraphics[width=5.45cm]{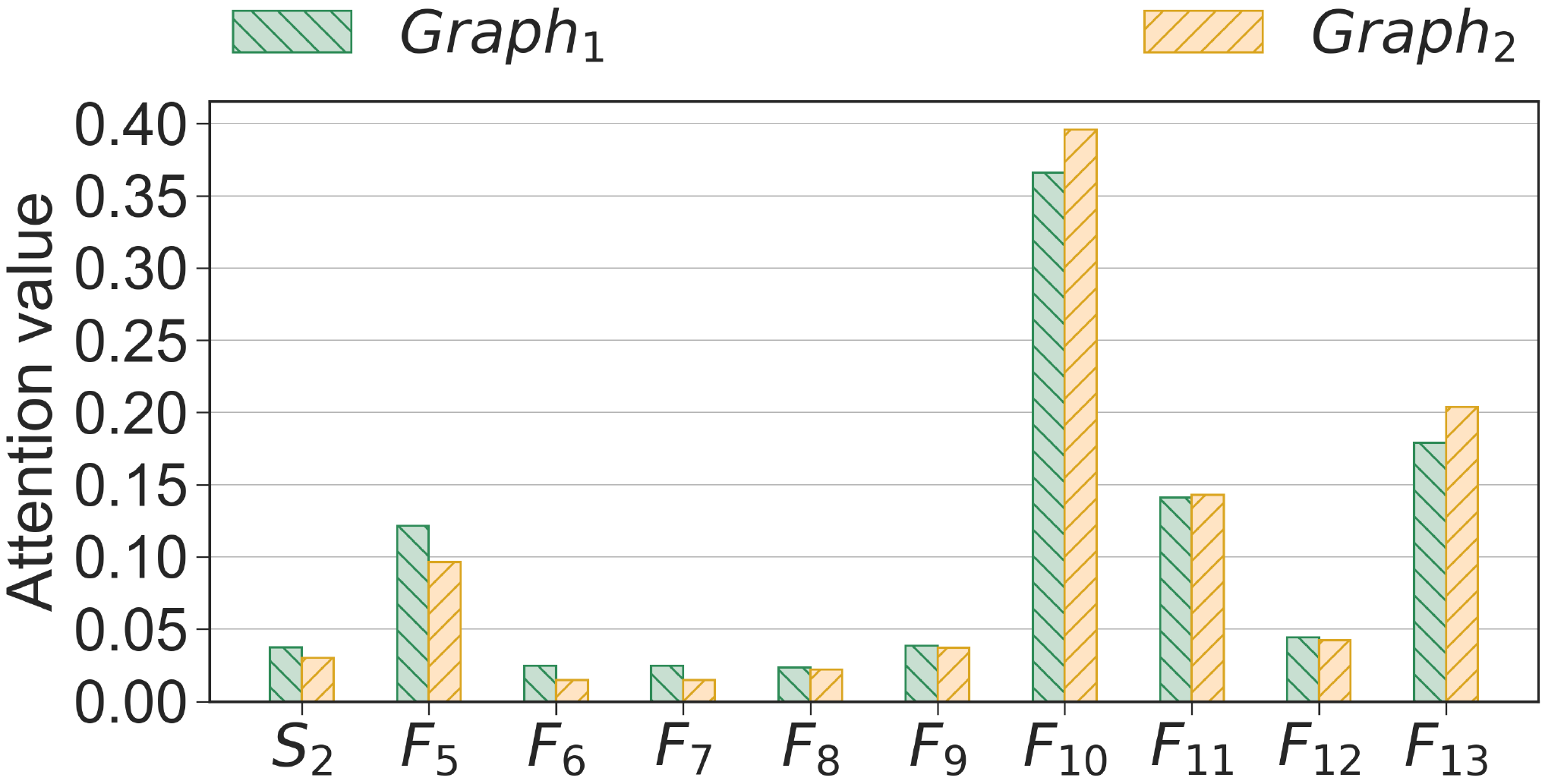}
\end{minipage}
\label{5c}
}
\subfigure[Semantic attention]{ 
\begin{minipage}[t]{0.31\linewidth}
\centering
\includegraphics[width=2.9cm]{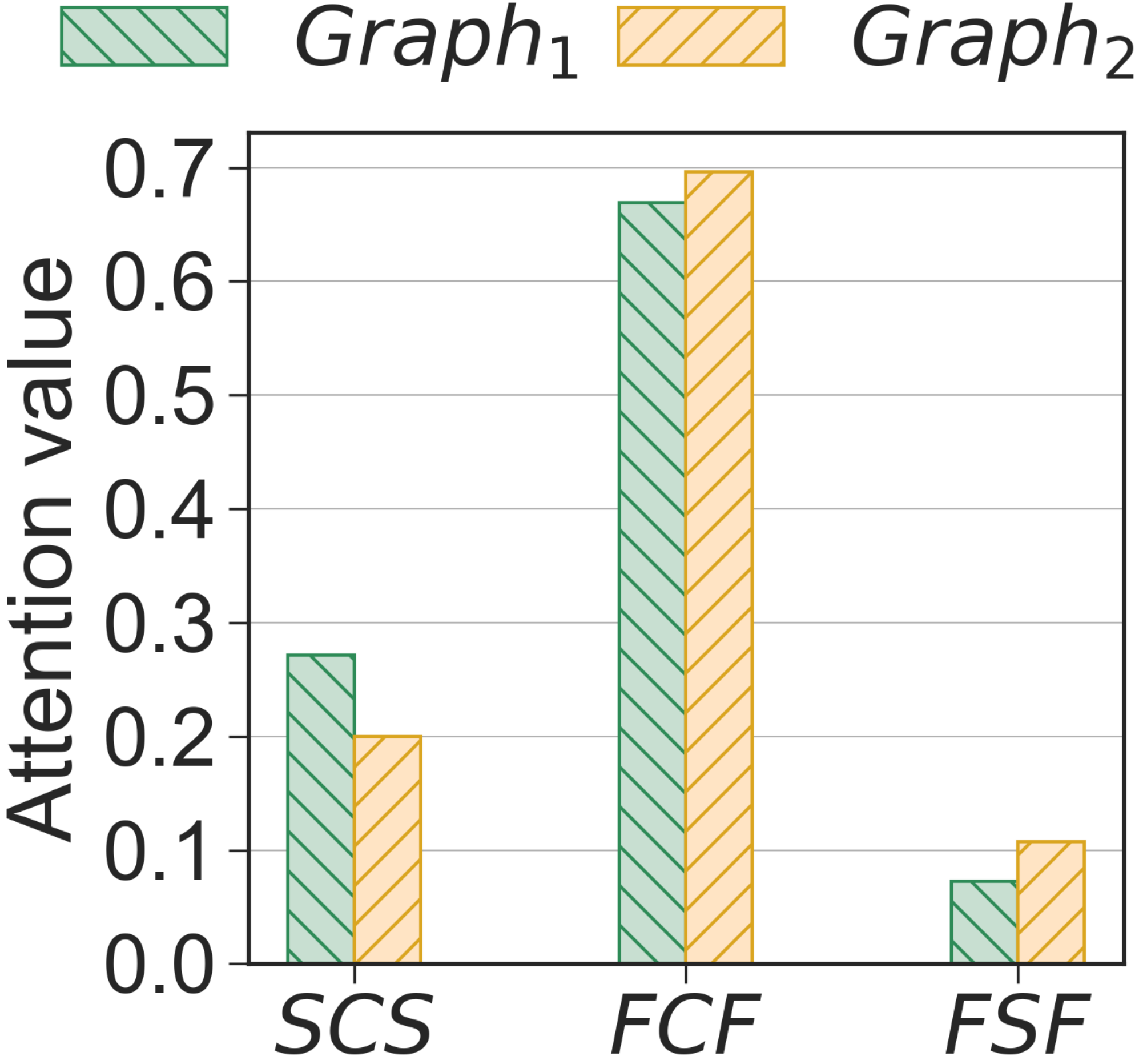}
\end{minipage}
\label{5d}
}
\centering
\subfigure[Graph attention on partial nodes $C \sim F_{13}$]{ 
\begin{minipage}[t]{1.0\linewidth}
\centering
\includegraphics[width=8.53cm]{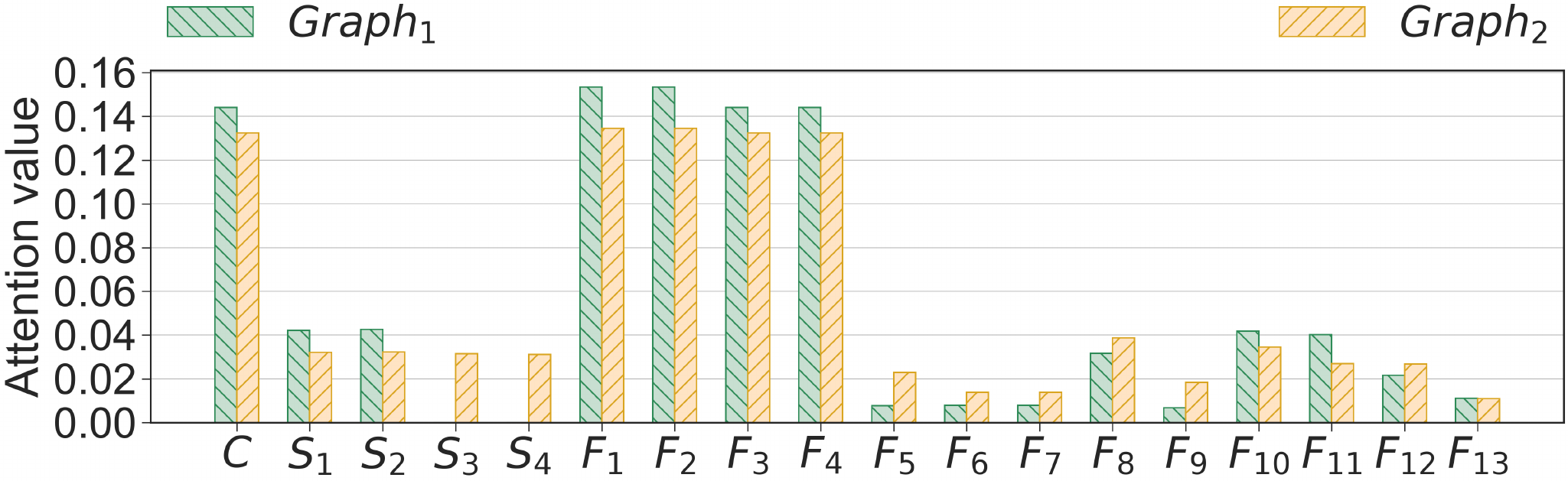}
\end{minipage}
\label{5e}
}
\centering
\vspace{-0.1cm}
\caption{A case study of hierarchical attention to help learn the similarity between two addresses’ meta-information.}
\label{fig5}
\end{figure}

To implement IPv6 client address correlation attacks, a salient property of \textsc{SiamHAN} is the incorporation of the hierarchical mechanism, which takes the importance of similar meta-information from two client addresses to help distance metric. Figure \ref{fig5} shows a case of two addresses' knowledge graphs and the hierarchical attention on partial nodes. In this setting, the two client nodes $C$ \footnote{The node attributes of client node $C$ are 2001:db8:3999::d05b:e903:1e77 and 2001:db8:880b::d05b:e903:1e77 in the two knowledge graphs.} link to 2 and 4 server nodes $S$ respectively, where $S_2$ is the common destination address. $F_1 \sim F_4$ are the client fingerprint nodes link to $C$ and $F_5 \sim F_{13}$ are the server fingerprint nodes link to $S_2$. The corresponding feature of each node label is shown in Table \ref{tab2}. 

\textbf{Analysis of Node-level Attention.} Node-level attention focuses on significantly similar meta-information in each meta-path-based neighbor between two graphs, a local view on each node to learn the semantic-specific embeddings. For instance, the $SCS$ meta-path-based neighbor attention of node $C$ is shown in Figure~\ref{5a}. The high attention values of node $C$ and $S_2$ come from the constant IID in the address of $C$ and the common server address of $S_2$ between the two input graphs. In addition, among the $FCF$ meta-path-based neighbors in Figure \ref{5b}, $F_3$ reaches a high attention value except for $C$ due to the same cipher suits used. The other client fingerprint nodes obtain low attention because they are usually the same regardless of the correlated or not correlated addresses, which lacks distinction to learn the unique embedding for the correlation task. Finally, as an example of $FSF$ node-level attention shown in Figure \ref{5c}, $F_5, F_{10}, F_{11}, F_{13}$ correspond to the server fingerprints including SNI, issuer, subject, and flow count, respectively. Results indicate they are more important to contribute to the server service embeddings learning.

\textbf{Analysis of Semantic-level Attention.} Semantic-level attention aggregates the 3-type semantic-specific embeddings for each node through the importance weights, which denotes the importance of meta-paths for the correlation task. In Figure \ref{5d}, the $FCF$ meta-path reaches the most attention in both the two knowledge graphs. The result indicates that the semantic-specific embedding learned by user browser parameters  substantially affects the user correlation.

\textbf{Analysis of Graph-level Attention.} Graph-level attention provides a global view on the similar final embedding of all nodes in the two graphs. The weights of partial nodes $C \sim F_{13}$ are shown in Figure \ref{5e}. Among them, client node $C$ and client fingerprint nodes $F_1 \sim F_4$ obtain higher attention values than the server-related nodes. It indicates \textsc{SiamHAN} takes more attention to the strong correlation from the client meta-information while only keeping eyes on specific server meta-information due to the complexity of user activity. Finally, \textsc{SiamHAN} could effectively leverage the meta-information integrated from address-based and traffic characteristic correlation to perform correlation attacks.

\begin{figure}[tb]
\centerline{\includegraphics[width=0.45\textwidth]{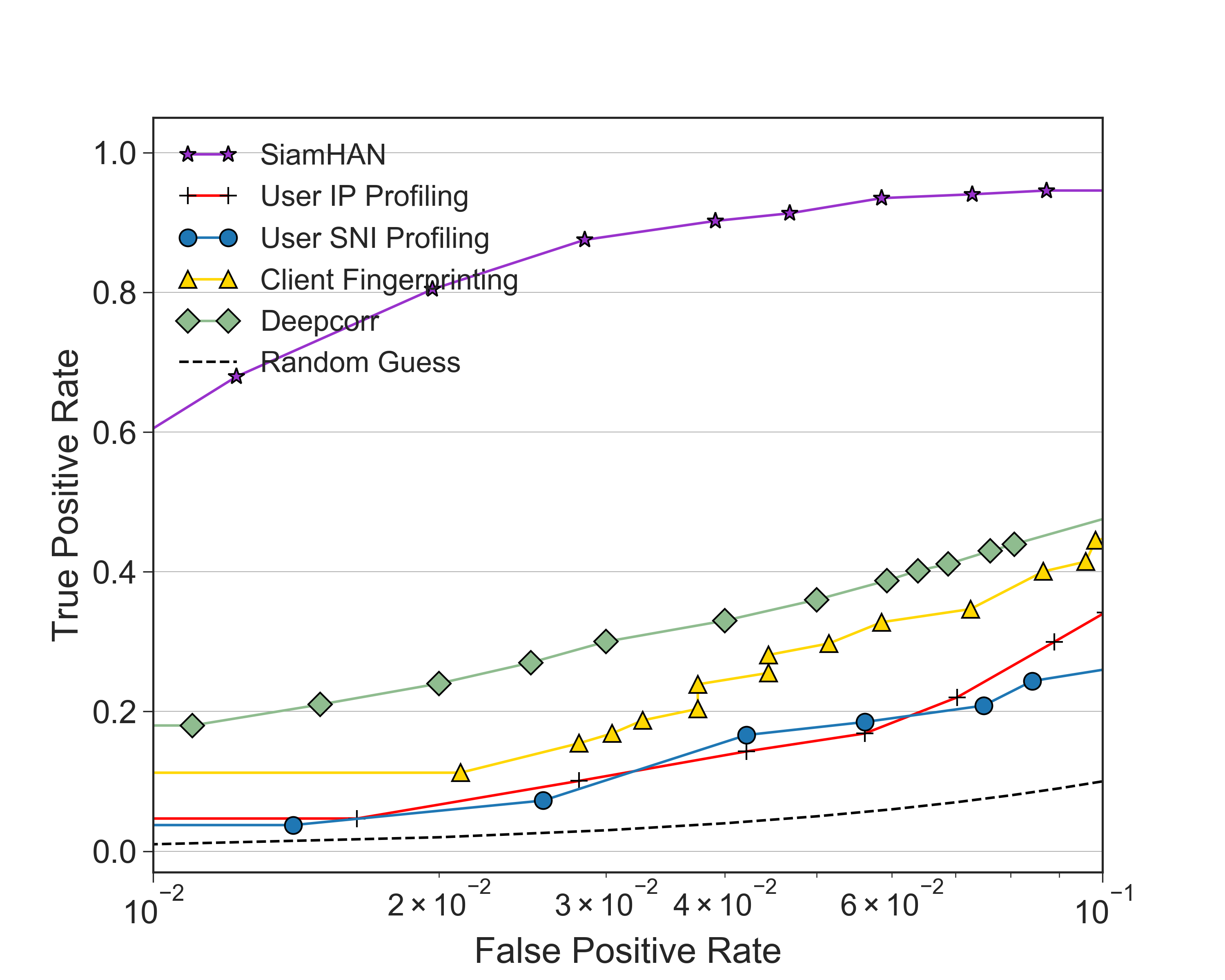}}
\caption{The performance of all baselines compared to \textsc{SiamHAN} based on the 5-month time-based split dataset.}
\label{fig6}
\end{figure}

\begin{figure}[!h]
\centerline{\includegraphics[width=0.45\textwidth]{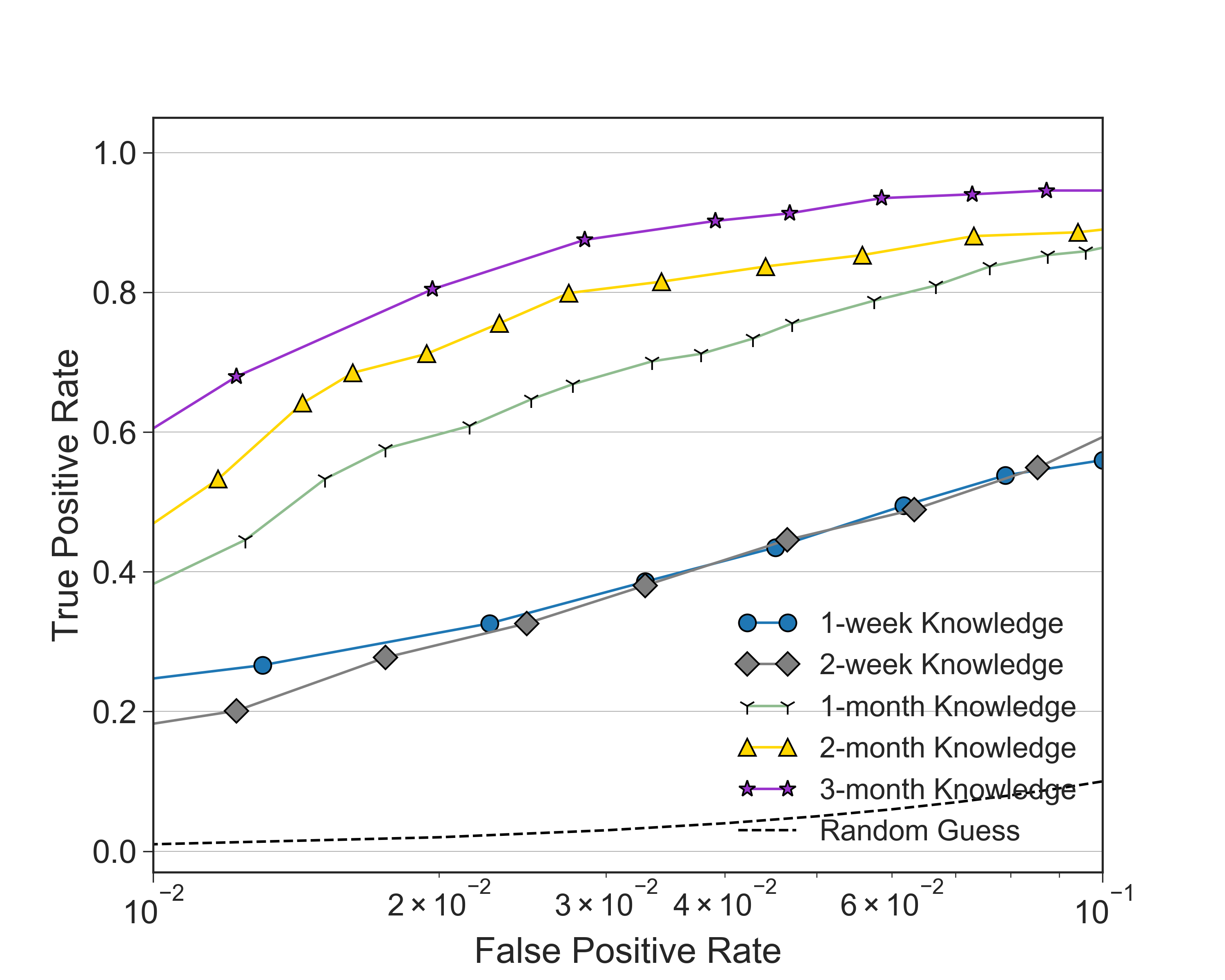}}
\caption{\textsc{SiamHAN}'s performance on different background knowledge volume $\kappa_t$ formed by wiretapping times $t$.}
\label{fig7}
\end{figure}

\subsection{Address Correlation}\label{sec6.3}
To explore the effectiveness of IPv6 address correlation attacks, we first measure the correlation performance of arbitrary address pairs performed by \textsc{SiamHAN}. An adversary could conduct correlation attacks on arbitrary pairwise addresses in this experimental setting based on the background knowledge $\kappa_t$. We comprehensively evaluate SiamHAN's performance on pairwise addresses correlation tasks by constructing training pair samples and test pair samples.

\begin{figure}[tb]
\centerline{\includegraphics[width=0.45\textwidth]{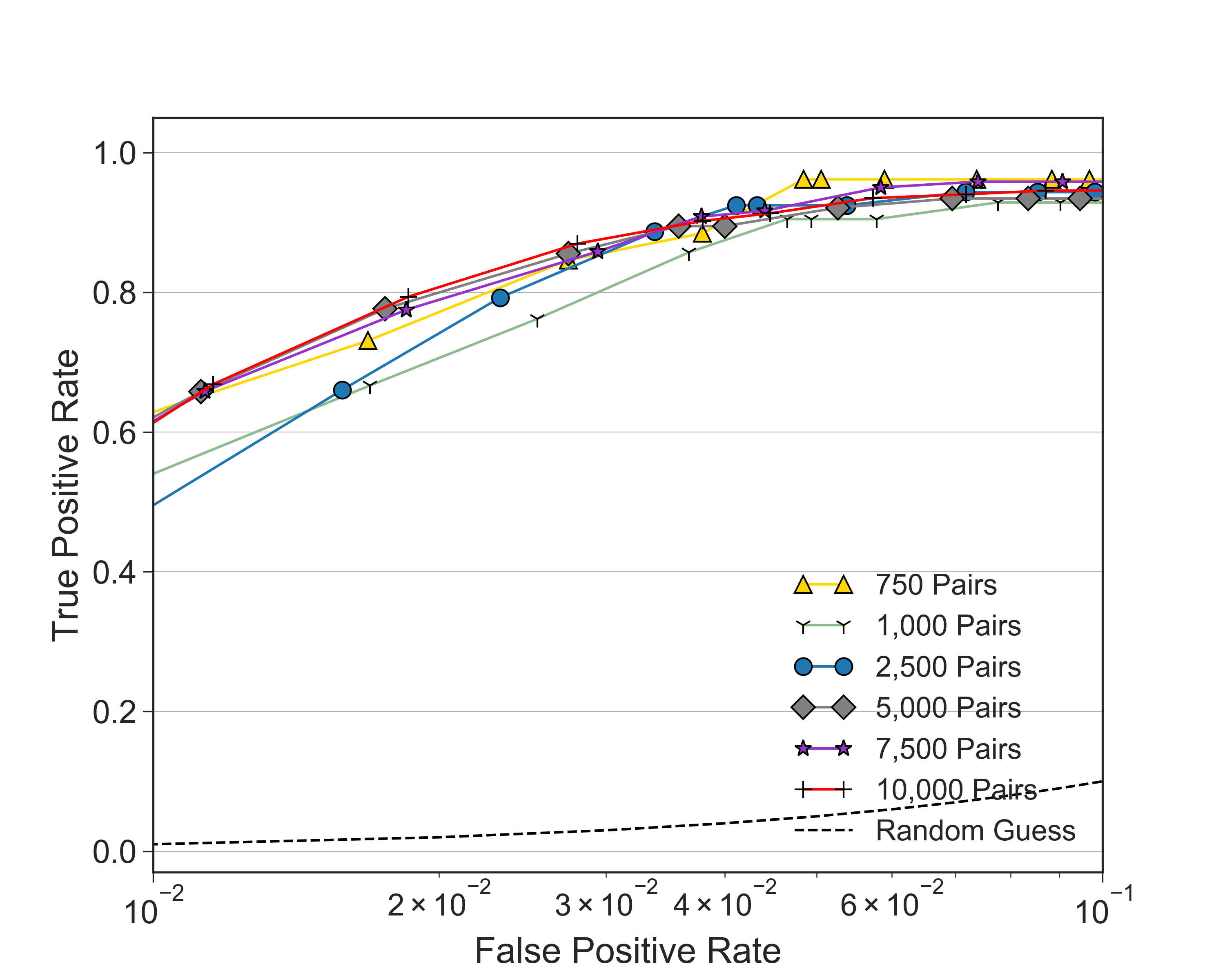}}
\caption{\textsc{SiamHAN}’s performance is consistent regardless of the test size in the time-based split dataset.}
\label{fig8}
\end{figure}

\begin{figure}[!h]
\centerline{\includegraphics[width=0.45\textwidth]{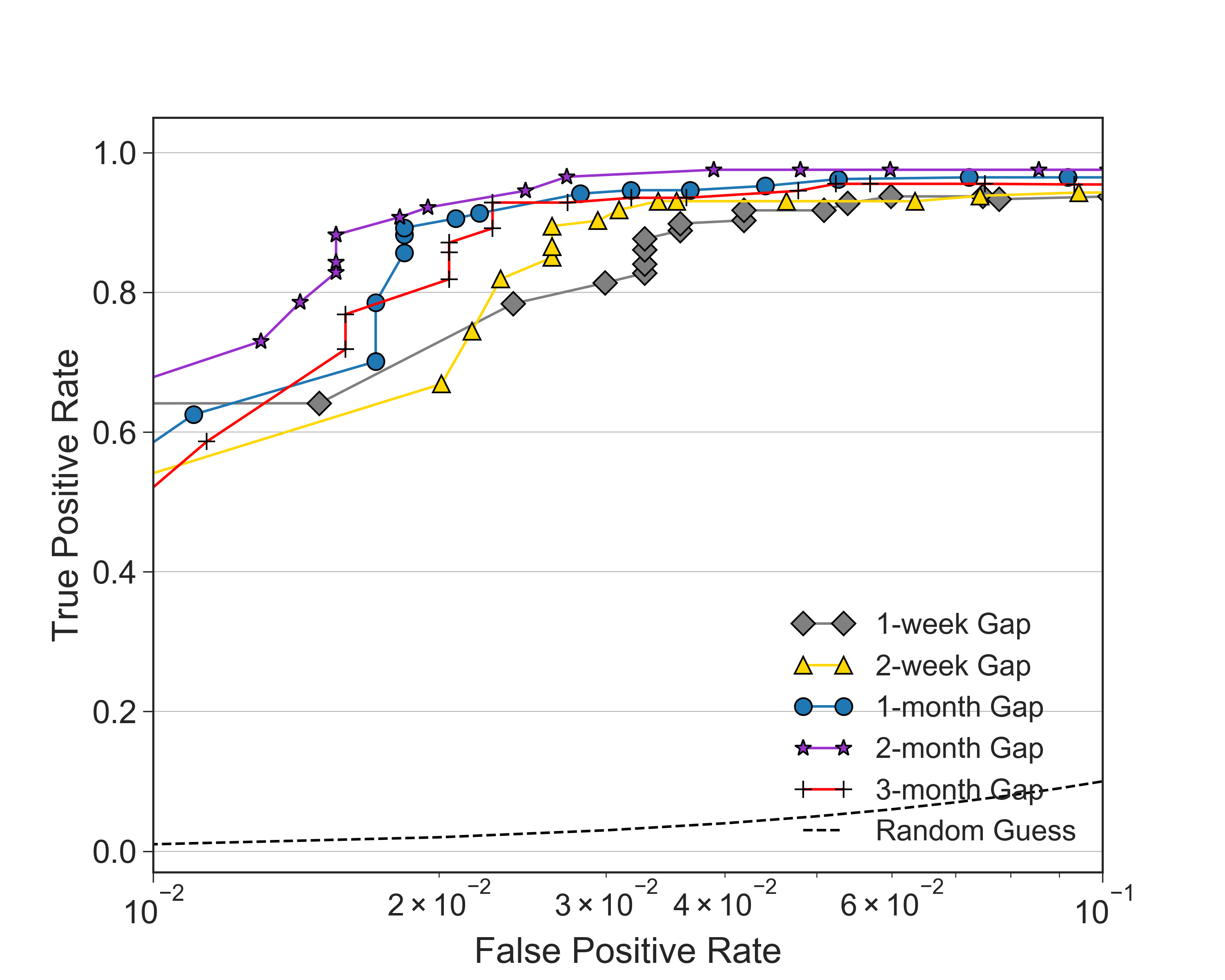}}
\caption{\textsc{SiamHAN}'s performance on different time gaps between training and test dataset for timeliness evaluation.}
\label{fig9}
\end{figure}

\textbf{Correlation Performance.} As a first look at the correlation performance, we train the attack model on the dataset with a time-based split setting. Figure \ref{fig6} compares the ROC curve of \textsc{SiamHAN} to other systems on the dataset. As can be seen, \textsc{SiamHAN} significantly outperforms the prior correlation algorithms with a wide gap between the ROC curve of \textsc{SiamHAN} and other approaches. For instance, for a target FPR = $4\times 10^{-2}$, while \textsc{SiamHAN} achieves a TPR of 0.90, all baselines provide TPRs less than 0.40. The drastic improvement comes from the fact that \textsc{SiamHAN} could model a correlation function tailored to pairwise client addresses with moderate learning on the knowledge of addresses. Since the test dataset contains addresses from not seen users in the training set, \textsc{SiamHAN} keeps a strong practical ability to correlate unknown addresses on the open-world dataset. 

\textbf{Adversary’s Background Knowledge.} In an IPv6 address correlation attack, the adversary's background knowledge $\kappa_t$ is essential to build the knowledge graph for each intercepted address. In Figure \ref{fig7}, we measure the impact of knowledge volume on \textsc{SiamHAN}'s performance with different wiretapping times $t$. This experiment set the training addresses with different duration of knowledge to build the training graph samples. Results indicate that a less than 2-week wiretapping time can not perform a strong enough correlation ability due to the weak knowledge volume. For a target FPR = $10^{-1}$, the adversary is only required 1-month monitoring to provide a TPR of 0.85, which could effectively correlate arbitrary address pairs with 90\% accuracy. \textsc{SiamHAN}'s performance is positively correlated with the volume of the adversary's background knowledge on the training set. 

\textbf{Robustness of Test Users.} On the consideration of \textsc{SiamHAN}'s practicality, we also show the correlation performance on the different sizes of the test dataset. Figure~\ref{fig8} presents the ROC curve results on test datasets with different numbers of sample pairs. The results are consistent for different numbers of addresses being correlated. It suggests the robustness of \textsc{SiamHAN} on the diverse user data. \textsc{SiamHAN} could provide stable correlation performance even when applied on significantly larger datasets of intercepted addresses, e.g., on the traffic collected by a large malicious IXP. 

\textbf{Timeliness.} Since the traffic characteristics of IPv6 users change over time, the deep learning-based algorithm requires timeliness evaluation to conduct a long-term reliable performance. Figure \ref{fig9} compares the results with different time gaps between training and test. In this experimental setting, we train \textsc{SiamHAN} on the dataset collected in the first month and test the pre-trained model on the same 1-month background knowledge dataset after different time gaps. The results indicate that \textsc{SiamHAN}'s performance does not degrade with the long-time gap. For a target FPR = $10^{-1}$, under all time gaps, \textsc{SiamHAN} provides TPRs more than 0.95, demonstrating the continuous effectiveness of the correlation attack model.

\begin{figure}[tbp]
\centerline{\includegraphics[width=0.45\textwidth]{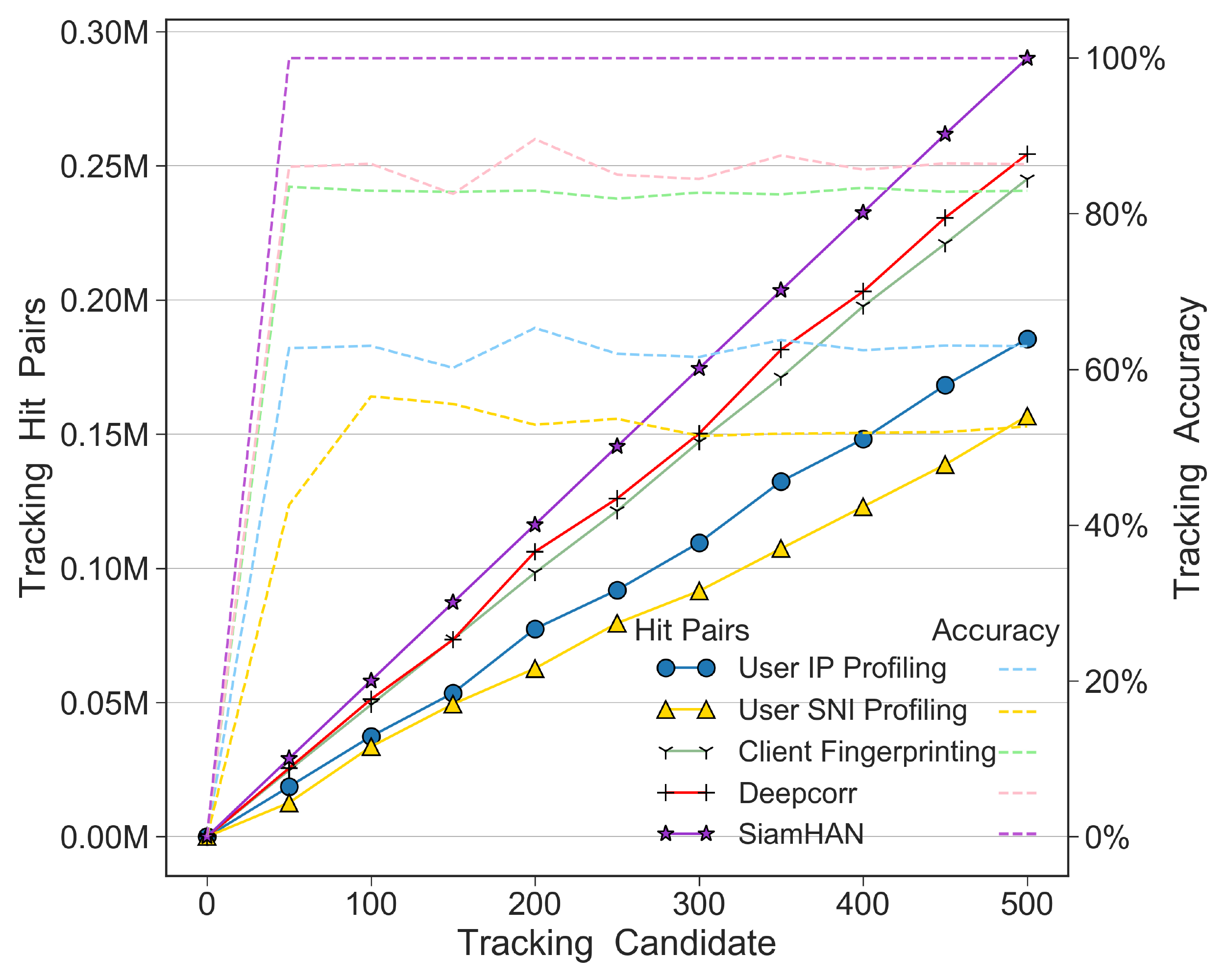}}
\caption{The tracking performance of all baselines compared to \textsc{SIamHAN} on the 5-month time-based split dataset with different sizes of tracking candidate $|S|$.}
\label{fig10}
\end{figure}

\subsection{User Tracking}
After obtaining the pre-trained attack model, the IPv6 address correlation attack could be applied to long-term user correlation tasks. We introduce user tracking, the first challenging task to sustainedly search target IPv6 users under the large-scale TLS encrypted traffic.

\textbf{Tracking Algorithm.} Given a pre-trained \textsc{SiamHAN}, based on the background knowledge, an adversary could conduct long-time user tracking by searching all addresses correlated to the address sample of target users. For a determined tracking candidate set $S$ that contains the one client address for each target user, the adversary is required to combine each candidate $S_i$ with each test address $T_j$ in the test dataset $T$ and build the pairs of their knowledge graphs $\langle S_i,\, T_j \rangle$ as \textsc{SiamHAN}'s inputs, where $i \leq |S|$ and $j \leq |T|$. Appendix \ref{app2} shows the detail of the tracking algorithm.

\textbf{Tracking Performance.} To implement IPv6 user tracking, we train \textsc{SiamHAN} and all compared baselines on the time-based split training set and select target user addresses from the test users to measure the tracking performance on the test dataset. Figure \ref{fig10} indicates the tracking performance of all baselines and \textsc{SiamHAN} in the user tracking task. As can be seen, \textsc{SiamHAN} could correctly identify 1.10$\sim$1.19 times more address pairs associated or non-associated with the target user samples than the state-of-the-art correlation system Deepcorr. \textsc{SiamHAN} outperforms existing correlation techniques with 99\% accuracy compared to Deepcorr's 85\% accuracy on the user tracking task. 

\begin{figure}[tbp]
\centerline{\includegraphics[width=0.45\textwidth]{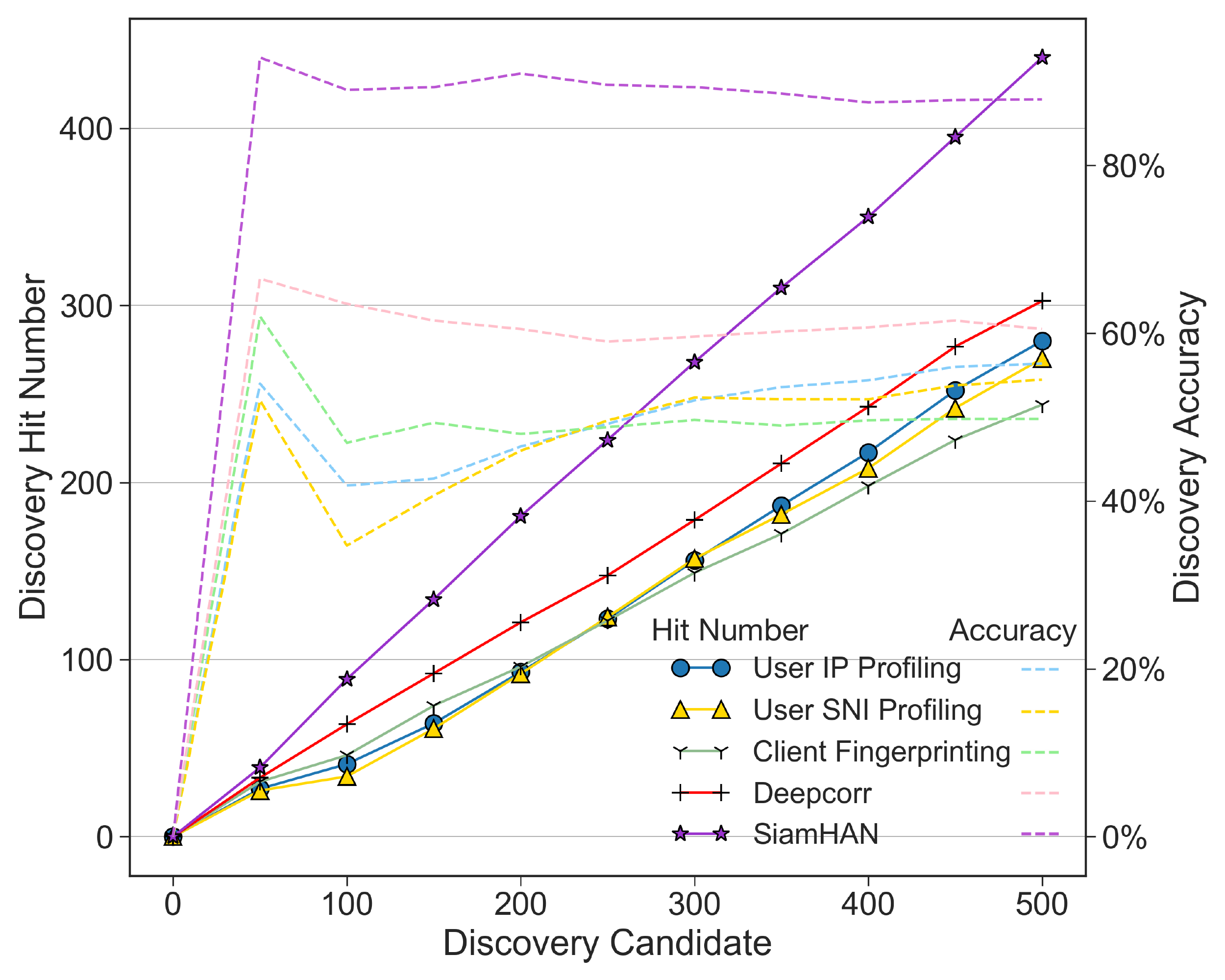}}
\caption{The discovery performance of all baselines compared to \textsc{SIamHAN} on the 5-month time-based split dataset with the different number of test users (discovery candidate).}
\label{fig11}
\end{figure}

\begin{table*}[t]
\caption{Ablation study on the 5-month time-based split dataset with all 3 experimental tasks.}
\begin{center}
\begin{tabular}{ccccccccc}
\toprule
\multirow{2}{*}{\textbf{Model}} && \multicolumn{2}{c}{\textbf{Address Correlation}} && \multicolumn{2}{c}{\textbf{User Tracking}} && \textbf{User Discovery}\\
\cline{3-4}\cline{6-7}\cline{9-9}
&&AUC&Accuracy&&AUC&Tracking Accuracy&&Discovery Accuracy\\
\midrule
Siamese GraphSAGE &&0.942&0.908&&0.933&0.913&&0.847\\
Siamese GAT &&0.955&0.922&&0.960&0.958&&0.875\\
\hline
\textsc{SiamHAN} - Client&&0.906&0.902&&0.944&0.930&&0.842\\
\textsc{SiamHAN} - Server&&0.920&0.911&&0.953&0.949&&0.863\\
\hline
\textsc{SiamHAN} - Node&&0.781&0.687&&0.847&0.880&&0.775\\
\textsc{SiamHAN} - Semantic&&0.912&0.906&&0.968&0.943&&0.865\\
\textsc{SiamHAN} - Graph&&0.909&0.879&&0.950&0.920&&0.839\\
\hline
\textsc{SiamHAN} - Classifier&&0.892&0.875&&0.887&0.892&&0.840\\
\hline
User IP Profiling&&0.785&0.711&&0.683&0.630&&0.564\\
User SNI Profiling&&0.777&0.693&&0.632&0.527&&0.545\\
Client Fingerprinting&&0.808&0.751&&0.794&0.829&&0.498\\
Deepcorr&&0.826&0.802&&0.819&0.855&&0.605\\
\hline
\textsc{SiamHAN}&&\textbf{0.966}&\textbf{0.932}&&\textbf{0.977}&\textbf{0.990}&&\textbf{0.880}\\
\bottomrule
\end{tabular}
\label{tab5}
\end{center}
\end{table*}

\subsection{User Discovery}
User discovery is the second challenging task applied by \textsc{SiamHAN}, which could obtain the address groups to discover unique IPv6 users on the large-scale encrypted traffic.

\textbf{Discovery Algorithm.} The adversary could construct the knowledge graph for each intercepted address and use a recursion algorithm to determine the unique users based on the adversary's background knowledge. The algorithm first selects an arbitrary address as the first user group. Then, in each iteration, the identified user group set is $G$, the algorithm calculates the average distance $\bar{D_i}$ between the new input address and each identified user group $G_i$'s addresses, where $i \leq |G|$. When all $\bar{D_i} > \eta$, we build a new user group $G_{|G|+1}$ for the current input address, while if some $\bar{D_i} \leq \eta$, we classify the input address into the user group with the closest distance. Appendix \ref{app3} shows the detail of the discovery algorithm.

\textbf{Discovery Performance.} In our experiments, to explore the performance of user discovery on the real-world 5-month dataset, we also train and test the model on the dataset with the time-based split setting. Figure \ref{fig11} indicates the discovery performance of all baselines and \textsc{SiamHAN} in the user discovery task. Results show a massive gap between \textsc{SiamHAN} and previous correlation approaches. For instance, \textsc{SiamHAN} provides a discovery accuracy of 88\% compared to 60\% by the state-of-the-art system Deepcorr using the same setting, which comes from \textsc{SiamHAN}'s $1.40 \sim1.54$ times more hit than Deepcorr. The significantly high accuracy of \textsc{SiamHAN} ensures the practicality to discover active IPv6 users in the wild traffic. 

\subsection{Ablation Study}
In addition to showing the experiments on the specific tasks, we present the ablation study experiments by evaluating the variants of \textsc{SiamHAN} to indicate the model superiority sufficiently. Table \ref{tab5} shows all results of the ablation study.

\textbf{Embedding Learning with Other GNNs.} We further investigate whether our attacks can apply to other GNNs with Siamese Networks. Concretely, we focus on GraphSAGE \cite{HamiltonYL17} and GAT \cite{VelickovicCCRLB18}, which are also well-known for inductive learning like HAN. We implement Siamese GraphSAGE and Siamese GAT by replacing the heterogeneous graph attention component in \textsc{SiamHAN}. Results show GNNs' powerful performance to learn the final graph embedding for general IPv6 address correlation. The variants with other GNNs still outperform previous correlation systems in all three evaluation tasks. However, \textsc{SiamHAN} keeps an unreachable better attack performance than other GNNs due to the semantic learning from the communication in heterogeneous graphs.

\textbf{Effectiveness of Different Fingerprint Types.} Since we collect client and server fingerprints as the meta-information for traffic characteristic correlation, to observe the importance of the two types of fingerprints, we implement \textsc{SiamHAN} - Client and \textsc{SiamHAN} - Server for the fingerprints ablation study. The two variants respectively remove all client fingerprint nodes or all server fingerprint nodes when building the knowledge graphs. In the experimental results, \textsc{SiamHAN} - Client's performance is poorer than \textsc{SiamHAN} - Server. It indicates that client fingerprints contribute more to the correlation attack due to the more decrease of the performance when \textsc{SiamHAN} lacks client fingerprints. 

\textbf{Effectiveness of Different Level Attentions.} To explore the effectiveness of each level of attention in the hierarchical attention, we also present three attention variants including \textsc{SiamHAN} - Node, \textsc{SiamHAN} - Semantic, and \textsc{SiamHAN} - Graph. The three variants respectively remove node-level, semantic-level, or graph-level attention and assign the same importance to each neighbor, each meta-path, or each final node embedding in the graph. Compared to \textsc{SiamHAN}, the performance of \textsc{SiamHAN} - Node drastically degrades, which indicates that attention on each node's neighbors is essential for the correlation task. Among the three types of attention, semantic-level attention contributes the least. Every level of attention could provide effective improvement to finally lead to the significantly high accuracy of \textsc{SiamHAN}. 

\textbf{Effectiveness of Distance Metric.} To indicate the superiority of the distance function learned for correlation, we replace \textsc{SiamHAN}'s distance learning with a binary classifier to build the variant \textsc{SiamHAN} - Classifier, which is implemented by modifying the last layer of \textsc{SiamHAN} to be a fully connected layer with Softmax activation. Results indicate that \textsc{SiamHAN} still outperforms \textsc{SiamHAN} - Classifier with a great margin. Since distance learning provides a more precise description of the difference between the two knowledge graphs, the distance metric architecture outperforms the classifier on the correlation task. 

\subsection{Time Complexity}
For the user tracking task, the attack generally tracks only a limited number of target users in the network, thus the time complexity is $O(cN)$, where $c$ is the number of target users. For the user discovery task, the discovery algorithm could be simplified by only computing the correlation relationship between the test address and one of the clustered addresses in each iteration. Therefore, the time complexity is $O(kN)$, where $k$ is the number of the cluster category. It is the usual time complexity of most cluster algorithms like K-means~\cite{Jain08}. 

\section{Countermeasures}\label{sec7}
To mitigate IPv6 address correlation attacks, we discuss two possible countermeasures: (1) traffic obfuscation and (2) the mechanisms to reduce the adversary's attack chances.

\subsection{Traffic Obfuscation}
An intuitive countermeasure against IPv6 address correlation attacks is to obfuscate TLS traffic used by \textsc{SiamHAN}. We show four types of traffic obfuscation methods in Table~\ref{tab6}. C-Random and CF-Random respectively denote using random forged addresses or browser parameters to obtain random client nodes or combinations of random client fingerprint nodes. CF-Background and SF-Background are the methods of adding background traffic with different browsers or different online services. The background traffic volume is the same as the original traffic of each user. Results indicate that each single obfuscation method is not effective enough to defend against \textsc{SiamHAN} since the correlation attack focuses on multi-type meta-information to find the similarity. When given a combination to apply all four methods, \textsc{SiamHAN}'s accuracy significantly degrades due to the knowledge barrier, which indicates that defending against IPv6 address correlation attacks requires strict limitations for address-based and traffic characteristic correlation.

\subsection{Attack Chance Reduction}
Another countermeasure against IPv6 address correlation attacks is reducing an adversary’s attack chances: (1) Since the adversary requires long-term monitoring to form the background knowledge, IPv6 users could escape the measurement on malicious IXPs by using proxies or Tor system. (2) To protect the meta-information exposed in the TLS handshake, applications like encrypted VPN could fundamentally render the attack impracticable. (3) We argue that designing address-user relation protection techniques like NAT is a promising avenue to defend against IPv6 address correlation attacks. 

\begin{table}[tbp]
\caption{\textsc{SiamHAN}'s accuracy with traffic obfuscation methods on the 5-month dataset with the time-based split.}
\vspace{-0.3cm}
\begin{center}
\begin{tabular}{cccc}
\toprule
\textbf{Obfuscation} & \textbf{Address} & \textbf{User} & \textbf{User}\\
\textbf{Method} & \textbf{Correlation} & \textbf{Tracking} & \textbf{Discovery}\\
\midrule
C-Random&0.855&0.905&0.808\\
CF-Random&0.878&0.897&0.810\\
CF-Background&0.871&0.922&0.823\\
SF-Background&0.893&0.910&0.830\\
\hline
Combination&0.705&0.769&0.643\\
\bottomrule
\end{tabular}
\label{tab6}
\end{center}
\end{table}

\section{Conclusion}\label{sec8}
In this work, we explore the implementation of user activity correlation on IPv6 networks. We propose IPv6 address correlation attacks, which leverage an attack model \textsc{SiamHAN} to learn the correlation relationship between two arbitrary IPv6 addresses based on the background knowledge of TLS traffic. Through multi-level attention and metric learning on pairwise heterogeneous knowledge graphs, \textsc{SiamHAN} could perform strong address correlation even on the long-term correlation tasks, including user tracking and user discovery. Numerous experiments indicate that \textsc{SiamHAN}'s performance and practicality outperform state-of-the-art algorithms by significant margins. 
We hope that our work demonstrates the serious threat of IPv6 address correlation attacks and calls for effective countermeasures deployed by the IPv6 community.

\section*{Acknowledgment}
We sincerely appreciate the shepherding from Matthew Wright and the writing help from Xinyu Xing. We would also like to thank the anonymous reviewers for their constructive comments and input to improve our paper. This work is supported by The National Key Research and Development Program of China (No.2020YFB1006100, No. 2018YFB1800200 and No. 2020YFE0200500) and Key research and Development Program for Guangdong Province under grant No. 2019B010137003.

%
%
%

\bibliographystyle{plain}
\bibliography{\jobname,mybibfile}

\section*{Appendix}
\appendix
\renewcommand{\appendixname}{Appendix~\Alph{section}}

\section{Analysis of Users without Plaintext Cookies}\label{app1}
The analysis of TLS users without plaintext cookies is shown in Table \ref{tab7}. Comparing with Table \ref{tab3}, results indicate that the source and online habits of the users with plaintext cookies are similar to the users without plaintext cookies.

\begin{table}[htb]
\caption{The analysis of TLS users without plaintext cookies with 2 dimensions including the top ASes of client addresses and the top SNI accessed by users.}
\begin{center}
\begin{tabular}{ll|ll}
\toprule
\textbf{AS Name}  & \textbf{\%Hits} & \textbf{SNI} & \textbf{\%Hits} \\
\midrule
CSTNET&75.2\%&*.google.com&18.3\%\\
China Unicom&10.3\%&*.adobe.com&14.6\%\\
CNGI-CERNET2&5.1\%&*.microsoft.com&13.2\%\\
CERNET&3.0\%&*.facebook.com&7.8\%\\
Reliance Jio&1.3\%&*.cloudflare.com&4.3\%\\
TSINGHUA6&0.7\%&*.icloud.com&4.0\%\\
Cloudflare&0.6\%&*.exoclick.com&2.8\%\\
PKU6-CERNET2&0.5\%&*.macromedia.com&2.3\%\\
ZZU6-CERNET2&0.5\%&*.flurry.com&1.4\%\\
\bottomrule
\end{tabular}
\label{tab7}
\end{center}
\vspace{-0.5cm}
\end{table}

\section{Tracking Algorithm}\label{app2}
The detailed tracking algorithm is shown in Algorithm \ref{alg1}, which exploits \textsc{SiamHAN} to search for addresses in the test set that belong to the same user as the tracking candidates.

\begin{algorithm}[htbp]
\caption{The tracking algorithm applied by \textsc{SiamHAN}} 
\label{alg1} 
\begin{algorithmic}[1] 
\REQUIRE Pre-trained \textsc{SiamHAN} $\rho$; Tracking candidate set~$S$; Test address set $T$; Background knowledge $\kappa_t$.
\ENSURE Address sets  $T_{S_i}$ link to the same user with each $S_i$
\FOR{$S_i$ in tracking candidate set $S$, where $i \leq |S|$}
\STATE Initialize target address set $T_{S_i}=\{\}$
\FOR{$T_j$ in test address set $T$, where $j \leq |T|$} 
\STATE Build pairwise knowledge graphs for $\langle S_i,\, T_j \rangle$
\STATE Test relationship $R$ of $\langle S_i,\, T_j \rangle$ using pre-trained $\rho$
\ENDFOR
\STATE Append $T_j$ in address set $T_{S_i}$ if relationship $R=1$
\ENDFOR 
\RETURN $T_{S_i}$ for each $S_i$
\end{algorithmic} 
\end{algorithm}

\section{Discovery Algorithm}\label{app3}
The detailed discovery algorithm is shown in Algorithm \ref{alg2}, which exploits \textsc{SiamHAN} to discover the unique users in the discovery candidate set.

\begin{algorithm}[htbp] 
\caption{The discovery algorithm applied by \textsc{SiamHAN}} 
\label{alg2} 
\begin{algorithmic}[1] 
\REQUIRE Pre-trained \textsc{SiamHAN} $\rho$; Discovery candidate set~$S$; Background knowledge $\kappa_t$; Task threshold $\eta$. 
\ENSURE User groups $G$ under the discovery candidate set $S$
\STATE Build knowledge graphs for each $S_i$
\STATE Initialize user group set $G = \{ G_1\}$
\STATE Initialize $S_1$ into the first user group $G_1$ 
\FOR{$S_i$ in discovery candidate set $S$, where $1 < i \leq |S|$}
\FOR{$G_k$ in user group set $G$} 
\FOR{Address $S_j$ in group $G_k$, where $j \leq |G_k|$} 
\STATE Calculate distance $D$ for $\langle S_i,\, S_j \rangle$ using $\rho$
\ENDFOR
\STATE Calculate average distance $\bar{D_k}$ for $S_i$ to $G_k$
\ENDFOR
\IF{All group average distance $\bar{D_k} > \eta$}
\STATE Initialize a new user group $G_{|G|+1}$ into $G$
\STATE Initialize $S_i$ into the new user group $G_{|G|+1}$ 
\ELSE
\STATE Classify $S_i$ into $G_k$ with the minimum $\bar{D_k}$
\ENDIF
\ENDFOR 
\RETURN User group set $G$
\end{algorithmic} 
\end{algorithm}
    

\end{document}